\documentclass[apj,iop]{emulateapj}
\usepackage{natbib}
\usepackage{graphicx}
\usepackage{amsmath}

\usepackage[bookmarks=false, colorlinks=true, citecolor=blue, linkcolor=blue, urlcolor=blue]{hyperref}

\newcommand{\Herschel}{\textit{Herschel}}
\newcommand{\Spitzer}{\textit{Spitzer}}
\newcommand{\Msun}{\hbox{$M_{\odot}$}}
\newcommand{\Lsun}{\hbox{$L_{\odot}$}}
\newcommand{\kms}{\hbox{\hbox{km}\,\hbox{s}$^{-1}$}}
\newcommand{\Lir}{\hbox{$L_{\rm IR}$}}
\newcommand{\FTS}{\hbox{SPIRE\slash FTS}}
\newcommand{\NCOav}{<\!N_{CO}\!>}

\newcommand{\Hii}{\hbox{\ion{H}{2}}}
\newcommand{\Nii}{\hbox{[\ion{N}{2}]1461\,GHz}}
\newcommand{\Cia}{\hbox{[\ion{C}{1}]492\,GHz}}
\newcommand{\Cib}{\hbox{[\ion{C}{1}]809\,GHz}}
\def\CO#1{\newcount\coj
\coj=#1
\advance\coj -1
\hbox{CO\,$J=#1-\the\coj$}}

\long\def\symbolfootnote[#1]#2{\begingroup%
\def\thefootnote{\fnsymbol{footnote}}\footnote[#1]{#2}\endgroup} 

\shorttitle{Sub-millimeter Spectra of Local Active Galaxies}
\shortauthors{Pereira-Santaella et al.}

\begin{document}

\title{\Herschel\slash SPIRE Sub-millimeter Spectra of Local Active Galaxies$^{\star,\star\star}$}

\author{Miguel Pereira-Santaella\altaffilmark{1}, Luigi Spinoglio\altaffilmark{1}, Gemma Busquet\altaffilmark{1}, Christine~D. Wilson\altaffilmark{2}, Jason Glenn\altaffilmark{3}, Kate~G. Isaak\altaffilmark{4}, Julia Kamenetzky\altaffilmark{3}, Naseem Rangwala\altaffilmark{3}, Maximilien~R.~P. Schirm\altaffilmark{2}, Maarten Baes\altaffilmark{5}, Michael~J. Barlow\altaffilmark{6}, Alessandro Boselli\altaffilmark{7}, Asantha Cooray\altaffilmark{8}, Diane Cormier\altaffilmark{9}}

\altaffiltext{$^\star$}{\Herschel\ is an ESA space observatory with science instruments provided by European-led Principal Investigator consortia and with important participation from NASA.}
\altaffiltext{$^{\star\star}$}{Partly based on observations carried out with the IRAM 30\,m Telescope. IRAM is supported by INSU/CNRS (France), MPG (Germany) and IGN (Spain).}
\altaffiltext{1}{Istituto di Astrofisica e Planetologia Spaziali, INAF, Via Fosso del Cavaliere 100, I-00133 Roma, Italy; \email{miguel.pereira@ifsi-roma.inaf.it}}
\altaffiltext{2}{Dept. of Physics \& Astronomy, McMaster University, Hamilton, Ontario, L8S 4M1, Canada}
\altaffiltext{3}{Center for Astrophysics and Space Astronomy, 389-UCB, University of Colorado, Boulder, CO 80303, USA}
\altaffiltext{4}{ESA Astrophysics Missions Division, ESTEC, PO Box 299, 2200 AG Noordwijk, The Netherlands}
\altaffiltext{5}{Sterrenkundig Observatorium, Universiteit Gent, Krijgslaan 281 S9, B-9000 Gent, Belgium}
\altaffiltext{6}{Department of Physics and Astronomy, University College London, Gower Street, London WC1E 6BT,UK}
\altaffiltext{7}{Laboratoire d'Astrophysique de Marseille - LAM, Universit\`e d'Aix-Marseille \& CNRS, UMR7326, 38 rue F. Joliot-Curie, 13388, Marseille Cedex 13, France}
\altaffiltext{8}{Center for Cosmology, Department of Physics and Astronomy, University of California, Irvine, CA 92697, USA}
\altaffiltext{9}{Laboratoire AIM, CEA/DSM-CNRS-Universit\`e Paris Diderot, Irfu/Service d'Astrophysique, CEA Saclay, F-91191, Gif-sur-Yvette, France}

\begin{abstract}
We present the sub-millimeter spectra from 450\,GHz to 1550\,GHz of eleven nearby active galaxies observed with the SPIRE Fourier Transform Spectrometer (\FTS) onboard \Herschel. We detect CO transitions from $J_{\rm up}=4$ to 12, as well as the two [\ion{C}{1}] fine structure lines at 492 and 809\,GHz and the \Nii\ line.
We used radiative transfer models to analyze the observed CO spectral line energy distributions (SLEDs). The FTS CO data were complemented with ground-based observations of the low-$J$ CO lines. We found that the warm molecular gas traced by the mid-$J$ CO transitions has similar physical conditions ($n_{\rm H_2}\!\sim$10$^{3.2}$--10$^{3.9}$\,cm$^{-3}$ and $T_{\rm kin}\!\sim$300--800\,K) in most of our galaxies. Furthermore, we found that this warm gas is likely producing the mid-IR rotational H$_2$ emission. We could not determine the specific heating mechanism of the warm gas, however it is possibly related to the star-formation activity in these galaxies.
Our modeling of the [\ion{C}{1}] emission suggests that it is produced in cold ($T_{\rm kin}< 30$\,K) and dense ($n_{\rm H_2}>10^3$\,cm$^{-3}$) molecular gas. Transitions of other molecules are often detected in our \FTS\ spectra. The HF $J=1-0$ transition at 1232\,GHz is detected in absorption in UGC~05101 and in emission in NGC~7130. In the latter, near-infrared pumping, chemical pumping, or collisional excitation with electrons are plausible excitation mechanisms likely related to the AGN of this galaxy.
In some galaxies few H$_2$O emission lines are present. Additionally, three OH$^+$ lines at 909, 971, and 1033\,GHz are identified in NGC~7130.

\end{abstract}


\section{Introduction}

The sub-millimeter (sub-mm) is one of the few astronomical spectral regions that has not been fully explored so far. In part, this is because ground-based sub-mm observations are difficult due to the low transparency of the atmosphere. In consequence, only a small number of bright sources have been observed in a few atmospheric windows (350, 450, and 850\,\micron).
The SPIRE Fourier transform spectrometer (\FTS; \citealt{Griffin2010SPIRE, Naylor2010, Swinyard2010}) onboard the \Herschel\ Space Observatory \citep{Pilbratt2010Herschel} fills most of this gap covering the spectral range between 450 and 1440\,GHz (210 and 670\,\micron) and provides an important benchmark for high-redshift studies with the Atacama Large Millimeter/submillimeter Array (ALMA).

Besides the continuum arising from cold dust, in the sub-mm range there are several molecular and atomic spectral lines that probe the different phases of the interstellar medium (ISM) in galaxies (e.g., \citealt{vanderWerf2010, Panuzzo2010, Rangwala2011}). In the \FTS\ spectral range the most prominent spectral feature is the CO ladder from $J_{\rm up}=4$ to 13. CO is a good tracer of the molecular gas since it is one of the most abundant molecules in the ISM. In particular, these mid-$J$ CO lines originate in warm molecular gas (their upper-level energies range from 55 to 500\,K above the ground state) that can be excited by ultraviolet (UV) photons in photon-dominated regions (PDRs; e.g., \citealt{Wolfire2010}), X-rays in X-ray dominated regions (XDRs; e.g., \citealt{Meijerink2006}) or by shocks  (e.g., \citealt{Flower2010}).
The analysis of the CO spectral line energy distribution (SLED), including low- and mid-$J$ CO transitions, is crucial for understanding the conditions in the cold and warm molecular gas and the dominant heating mechanism in galaxies \citep{Rangwala2011, Kamenetzky2012, Spinoglio2012}.

\begin{deluxetable*}{lccccccccccc}[ht!]
\tablewidth{0pt}
\tabletypesize{\small}
\tablecaption{Sample of Local Active Galaxies\label{tab:sample} }
\tablehead{ \colhead{Galaxy} & \colhead{R.A.}  & \colhead{Decl.} & \colhead{$cz^a$} & \colhead{Dist.$^b$} & \colhead{Spect. Class$^c$} & \colhead{Ref.$^d$} & \colhead{$\log {L_{\rm IR}}^e$} & \colhead{${L_{\rm X}}^f$} & \colhead{Ref.$^g$} \\[0.1ex]
            &   \colhead{(J2000.0)}           & \colhead{(J2000.0)}     &\colhead{(km\,s$^{-1}$)} & \colhead{(Mpc)} & & & \colhead{(\Lsun)} & (10$^{41}$\,erg\,s$^{-1}$) }
\startdata
NGC~1056 	& 	02 42 48.3  &   +28 34 27    & 1545  & 32.2 & \Hii   & 1 & 10.3 & \nodata & \nodata \\
UGC~05101 	& 	09 35 51.6  &   +61 21 11    & 11802 & 163  & LINER  & 1 & 12.0 & 75 & 7 \\
NGC~3227 	& 	10 23 30.6  &	+19 51 54    & 1157  & 14.4 & Sy1.5  & 2 & 9.9 & 1.9 & 8 \\
NGC~3982 	& 	11 56 28.1  &	+55 07 31    & 1109  & 20.6 & Sy1.9  & 2 & 10.0 & $<$0.025 & 9 \\
NGC~4051 	& 	12 03 09.6  &	+44 31 53    & 700   & 14.6 & Sy1.2  & 2 & 10.0 & 1.3 & 10 \\
NGC~4151 	& 	12 10 32.6  &	+39 24 21    & 995   & 12.1 & Sy1.5  & 2 & 9.8 & 8.2 & 10 \\
NGC~4388 	& 	12 25 46.7  &	+12 39 44    & 2524  & 18.2 & Sy1.9\slash HBLR & 2, 3 & 10.2 & 3.0 & 11 \\
IC~3639 	& 	12 40 52.8  &	 $-$36 45 21 & 3275  & 42.5 & Sy2\slash HBLR & 1, 4 & 10.8 & 0.17 & 12 \\ 
NGC~7130 	& 	21 48 19.5  &	 $-$34 57 04 & 4842  & 66.0 & Sy2 & 1 & 11.4 & 0.83 & 13 \\
NGC~7172 	& 	22 02 01.9  &	 $-$31 52 11 & 2603  & 33.9 & Sy2 & 5 & 10.4 & 29 & 10 \\
NGC~7582 	& 	23 18 23.5  &	 $-$42 22 14 & 1575  & 20.6 & \Hii\slash Sy1i & 1, 6& 10.8 & 2.0 & 14
\enddata
\tablecomments{$^{(a)}$ Heliocentric velocity from NASA\slash IPAC extragalactic database (NED). $^{(b)}$ Distance from NED. $^{(c)}$ Nuclear activity classification. Sy1i indicates that broad emission lines are observed in the near-IR. HBLR indicates the detection of a hidden broad line region. $^{(d)}$ Reference for the nuclear activity classification. $^{(e)}$ 8--1000\,\micron\ infrared luminosity from \citet{SandersRBGS} scaled to our adopted distance. $^{(f)}$ X-ray 2--10\,keV luminosity.  $^{(g)}$ Reference for the X-ray luminosity. }
\tablerefs{(1) \citet{Yuan2010}; (2) \citet{Ho1997}; (3) \citet{Young1996}; (4) \citet{Heisler1997}; (5) \citet{Lumsden2001}; (6) \citet{Reunanen2003}; (7) \citet{Imanishi2003b}; (8) \citet{Gondoin2003}; (9) \citet{Guainazzi2005}; (10) \citet{Brightman2011}; (11) \citet{Cappi2006}; (12) \citet{Guainazzi2005}; (13) \citet{Levenson2005}; (14) \citet{Piconcelli2007}.}
\end{deluxetable*}

In addition to the CO ladder, two [\ion{C}{1}] fine structure lines are present in the sub-mm range. PDR models predict that these lines originate in the transition region between C$^{+}$ and CO \citep{Tielens1985, Kaufman1999}. However, several mechanisms, such as turbulent diffusion, non chemical equilibrium, or intense X-ray radiation, can influence the C$^0$ distribution and abundance (\citealt{Papadopoulos2004} and references therein).

To date, there are relatively few published \FTS\ spectra of extragalactic objects \citep{vanderWerf2010,Panuzzo2010,Rangwala2011,Kamenetzky2012,Spinoglio2012,Meijerink2013}; in these, however, molecules such as H$_2$O, OH$^+$, and HF are often detected. In addition, other molecules like HCN, CH$^+$, and H$_2$O$^+$ have also been detected.

In this paper we present for the first time the complete sub-mm spectra of a sample of eleven active galaxies (Section \ref{sec:sample}). The sub-mm spectra are complemented by ground-based IRAM 30\,m observations of the \CO1 and $J = 2-1$ transitions in three galaxies.
The SPIRE and IRAM 30\,m observations and data reduction are described in Section \ref{s:observations}. We use the escape probability approximation to model the CO and C$^0$ emission and study how they trace the warm and cold molecular gas in these galaxies (Sections \ref{s:rad_transfer} and \ref{s:neutral_carbon}). In two galaxies we detect the HF $J = 1-0$ transition, one in emission and one in absorption. We discuss several possible origins for this line in Section \ref{s:HF}. Finally, a brief discussion about the H$_2$O and OH$^+$ lines detected in a few galaxies is presented in Section \ref{s:other}. The [\ion{N}{2}] line at 205\,\micron\ (1461\,GHz) is one of the brightest far-IR lines in galaxies (e.g., \citealt{Wrigh1991}), however it has not been systematically explored by previous observatories. In a forthcoming paper we will study the emission of the  [\ion{N}{2}]\,205\,\micron\ line and other far-IR atomic fine structure emission lines in our sample of galaxies.
Throughout this paper we assumed $H_{\rm 0} = 73$\,km\,s$^{-1}$\,Mpc$^{-1}$.

\section{Sample of Local Active Galaxies}\label{sec:sample}

We study a sample of eleven local ($d=12-160$\,Mpc) active galaxies drawn from the 12\,\micron\ Galaxy Sample \citep{Rush1993}. They were selected because of their high \textit{IRAS} 100\,\micron\ flux and to include the different nuclear activity classes in similar proportions (three Seyferts type 1.2 and 1.5, two type 1.9 Seyfert, three type 2 Seyfert, one low-ionization nuclear emission line region [LINER], and two \Hii\ galaxies). Their total infrared (IR) luminosities ($L_{\rm IR}$) range from 10$^{9.8}$ to 10$^{12.0}$\,\Lsun, and the median value is 10$^{10.3}$\,\Lsun. In our sample, only UGC~05101 and NGC~7130 are classified as luminous IR galaxies (LIRGs; $L_{\rm IR}>10^{11}$\,\Lsun), so in these galaxies, besides the AGN activity, intense star-formation is expected \citep{Petric2011, AAH2012a}. The galaxies in our sample are presented in Table \ref{tab:sample}.

\section{Observations and Data Reduction}\label{s:observations}

All the galaxies in our sample were observed with the \FTS\ and the SPIRE photometer onboard \Herschel. Additionally, we observed the \CO1\ and $J=2-1$ rotational transitions in UGC~05101, NGC~3227, and NGC~3982 with the IRAM 30\,m telescope. The SPIRE photometric data will be analyzed in detail in a forthcoming paper, but the data reduction is summarized here because the images are used to calculate the source-beam coupling factor and the scaling factor for the flux calibration of the \FTS\ spectra.

\begin{deluxetable}{lccccccccccc}
\tablewidth{0pt}
\tabletypesize{\small}
\tablecaption{Log of \Herschel\ Observations\label{tab:journal}}
\tablehead{ \colhead{Galaxy} & \colhead{\FTS\ Spectrometer} & &\colhead{SPIRE Photometer}\\
& \colhead{Observation ID}  & & \colhead{Observation ID}}
\startdata
NGC~1056 	  & 1342204024 & & 1342226630 \\
UGC~05101 	  & 1342209278 & & 1342204962 \\
NGC~3227 	  & 1342209281 & & 1342197318 \\
NGC~3982 	  & 1342209277 & & 1342186862 \\
NGC~4051 	  & 1342209276 & & 1342210502 \\
NGC~4151 	  & 1342209852 & & 1342188588 \\
NGC~4388 	  & 1342210849 & & 1342211416 \\
IC~3639 	  & 1342213381 & & 1342202200 \\
NGC~7130 	  & 1342219565 & & 1342210527 \\
NGC~7172 	  & 1342219549 & & 1342209301 \\
NGC~7582 	  & 1342209280 & & 1342210529  
\enddata
\end{deluxetable}

\subsection{\FTS\ Spectroscopy}\label{ss:spectroscopy}

We collected \Herschel\ \FTS\ spectroscopic observations of eleven nearby active galaxies through a guaranteed time project (PI: L. Spinoglio) and a guaranteed time key project (PI: C.~D. Wilson). The integration times of the observations were $\sim$5100\,s, achieving a sensitivity of 0.2--0.3\,Jy in the continuum and a 3$\sigma$ detection limit of $\sim$1$\times10^{-14}$\,erg\,cm$^{-2}$\,s$^{-1}$ for an unresolved spectral line.

\begin{figure*}[!ht]
\centering
\includegraphics[width=1.8\columnwidth]{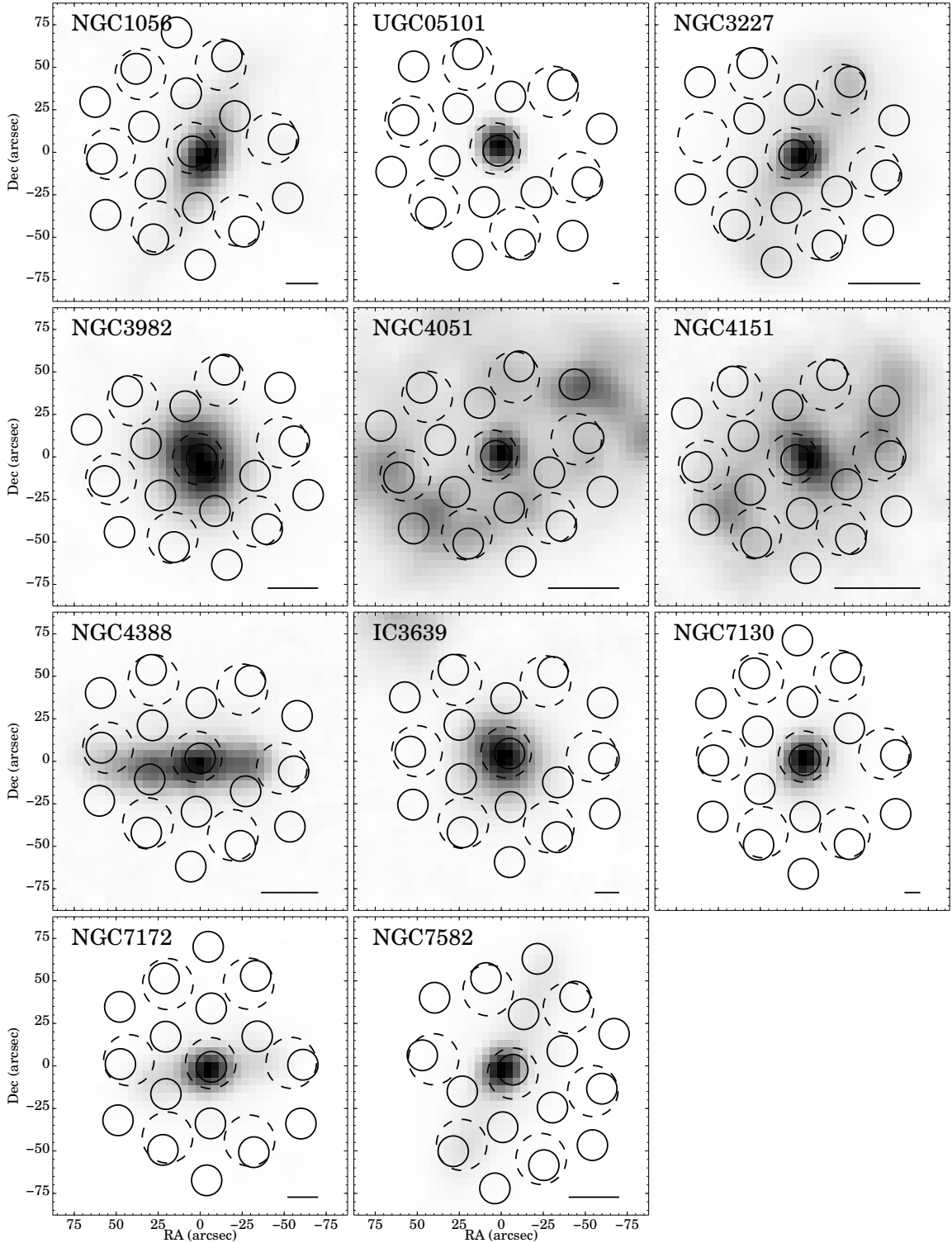}
\caption{Footprint of \FTS\ FOV plotted over the \Herschel\slash SPIRE 250\,\micron\ images of our sample of Seyfert galaxies. The solid and dashed circles mark the positions of the SSW and SLW bolometers, respectively. Only the functioning detectors in the unvignetted FOV are plotted. North is up and East to the left. The images are shown in a linear gray scale. The black line represents 3\,kpc at the distance of the galaxy. \label{fig:ftsfootprint}}
\end{figure*}

The \FTS\ has two bolometer arrays: the spectrometer short wavelength (SSW; 194--324\,\micron, 925--1545\,GHz) and the spectrometer long wavelength (SLW; 316--672\,\micron, 446--948\,GHz). The beam full-width half-maximum (FWHM) of the SSW bolometers is $\sim$18\arcsec, approximately constant with frequency. On the contrary, the beam FWHM of the SLW bolometers varies between $\sim$30\arcsec\ and $\sim$42\arcsec\ with a complicated dependence on frequency \citep{Swinyard2010}. They sample sparsely a field of view (FOV) of $\sim$2\arcmin, with the central bolometer of the arrays centered at the nuclei of the galaxies (see Figure \ref{fig:ftsfootprint}). At the distances of our galaxies 30\arcsec\ corresponds to physical scales of 2 to 20\,kpc.
All the galaxies were observed at high spectral resolution (unresolved line FWHM 1.45\,GHz, \hbox{$\nu \slash \Delta \nu$} $\sim 300-1000$) in the single pointing mode. The observation IDs are listed in Table \ref{tab:journal}.

We processed the raw data with the \Herschel\ interactive pipeline environment software (HIPE) version 9 \citep{Ott2010HIPE}. We used the standard reduction scripts \citep{Fulton2010}. These scripts process the raw \FTS\ timelines and produce a flux calibrated spectrum for each bolometer.
Briefly, first the bolometer timelines are corrected for instrumental effects and cosmic ray hits, and the interferograms are created. The baselines are removed and the phase of the interferograms corrected before a Fourier transform is applied to obtain the spectra. These spectra are dominated by the telescope (and instrument) thermal emission that must be removed to obtain the source spectrum.

The uncertainty in the telescope emission model used by HIPE is 1--2\,Jy in the continuum level (SPIRE Observer's Manual\footnote{Available at http://herschel.esac.esa.int/Docs/SPIRE/\\html/spire\_om.html}), which is comparable to the fluxes of our galaxies. Therefore, the residual telescope emission has to be subtracted before applying the point source correction to the spectra. The optical angular sizes of our galaxies are several arcmin, but the nuclear far-IR emission appears compact at the angular resolution of SPIRE (only UGC~05101 and NGC~7130 are almost point-like at the \FTS\ angular resolution, see Figure \ref{fig:ftsfootprint}). Moreover, far-IR emission is detected only in the central bolometer of the \FTS\ arrays (FWHM$\sim$18\arcsec\ and 30--40\arcsec\ for the SSW and SLW arrays, respectively). Thus, we estimated the residual telescope emission averaging the spectra of the off-axis bolometers. For some galaxies, some of the off-axis detectors contain extended galactic emission (weak continuum and \Nii\ emission). We excluded those detectors before averaging.

Next we applied the point source correction\footnote{\label{foot:h9}Taken from the HIPE version 9 calibration product \texttt{BeamParam\_HR\_unapod\_nominal\_20050222}. It provides the point source correction and the beam size as a function of frequency with a 0.3\,GHz sampling.} to the central bolometer spectra to obtain the flux calibrated spectra of the galactic nuclei. To correct for the large beam size variations in the SLW spectra we calculated the source-beam coupling factor, $\eta$. 
We define $\eta$ as the ratio between the flux measured with a beam of size $\theta$ and that measured with a 30\arcsec\ beam (the minimum SLW beam size).
Similar to \citet{Kamenetzky2012} and \citet{Panuzzo2010}, we convolved the 250\,\micron\ SPIRE maps using Gaussian profiles to obtain the coupling factor as a function of the beam size, $\eta(\theta)$. Then with the SLW beam size dependence on the frequency$^{\ref{foot:h9}}$, $\theta(\nu)$, we obtained $\eta(\nu)$ and we divided the SLW spectra by this function. The values of $\eta(\nu)$ for these galaxies vary between 1.0 and 1.4.
By applying this correction we assume that the spatial distributions of the emission lines in the \FTS\ spectra are comparable to that of the cold dust traced by the 250\,\micron\ emission.

Finally, we scaled the SSW and SLW spectra to match the SPIRE photometry (see Section \ref{ss:photometry}).
For the SSW we scaled the spectra to match the 250\,\micron\ (1200\,GHz) flux measured in a 18\arcsec\ diameter aperture.
Similarly, we scaled the SLW spectra to match the flux measured in the 350\,\micron\ (856\,GHz) image in a 30\arcsec\ diameter aperture. We used the 350\,\micron\ SPIRE image because for most of our sources the continuum at 500\,\micron\ (600\,GHz) is below the detection limit of the FTS data. Point source aperture corrections are applied to the SPIRE photometry (see Section \ref{ss:photometry}).
The scale factors vary between 0.85 and 1.30 for the SSW spectra and between 0.95 and 1.20 for the SLW spectra. For NGC\,1056, NGC\,4051, NGC\,4151, IC~3639, and NGC\,7172 the 350\,\micron\ continuum is not detected in the FTS spectra. Thus for these galaxies the uncertainty in the absolute flux of the emission lines in the SLW range is about 20\,\%\ larger.
The final flux calibrated spectra of the galaxies are shown in Figure \ref{fig:spectra}.

We measured the line intensities by fitting a sinc profile\footnote{The sinc function represents well the \FTS\ instrumental line shape (see the SPIRE Observer's Manual).} to the emission lines. To obtain the local continuum  we used a linear fit in a $\sim$10\,GHz range around the line position. When two lines were close in frequency (e.g., \CO{7} at 807\,GHz and \Cib) the two lines were fitted simultaneously.
In the \FTS\ spectra of our galaxies we find CO lines from $J_{\rm up}=4$ to $J_{\rm up}=12$, the \Cia, \Cib, and \Nii\ atomic lines, and some water lines. The HF $J=1-0$ rotational transition at 1232\,GHz is detected in two sources and in one source we detect three OH$^+$ transitions. The fluxes and 1$\sigma$ uncertainties of the detected lines and the 3$\sigma$ upper limits for the undetected lines are given in Tables \ref{tab:lines} and \ref{tab:lines_uncommon}.

\begin{deluxetable*}{lcccccccc}
\tablewidth{0pt}
\tabletypesize{\small}
\tablecaption{SPIRE\slash FTS Line Fluxes\label{tab:lines}}
\tablehead{
\colhead{Transition}& \colhead{\rm $\nu_{\rm rest}$}  &  \colhead{} & \multicolumn{6}{c}{\rm Fluxes (10$^{-15}$\,erg\,cm$^{-2}$\,s$^{-1}$)} \\
 \cline{4-9} \\
  &   (GHz)  &	   & NGC~1056 & UGC~05101 & NGC~3227 & NGC~3982 & NGC~4051 & NGC~4151 
}
\startdata
$^{12}$CO J = 4--3 & 461.041 &  &  $<$15.8 &  \nodata$^{a}$  & 29.0 $\pm$ 3.7 & 18.1 $\pm$ 2.7 &  $<$13.0 &  $<$9.5 \\
$^{12}$CO J = 5--4 & 576.268 &  & 10.1 $\pm$ 2.0 & 15.6 $\pm$ 3.8 & 33.4 $\pm$ 3.9 & 12.9 $\pm$ 2.2 &  $<$11.2 &  $<$8.0 \\
$^{12}$CO J = 6--5 & 691.473 &  & 6.0 $\pm$ 1.5 & 16.9 $\pm$ 2.3 & 25.2 $\pm$ 2.9 & 7.3 $\pm$ 1.6 & 9.6 $\pm$ 1.6 &  $<$4.1 \\
$^{12}$CO J = 7--6 & 806.652 &  & 6.1 $\pm$ 1.4 & 11.1 $\pm$ 1.8 & 18.1 $\pm$ 2.3 & 4.7 $\pm$ 1.3 & 7.3 $\pm$ 1.4 & 3.6 $\pm$ 1.1 \\
$^{12}$CO J = 8--7 & 921.800 &  &  $<$8.0 & 11.6 $\pm$ 2.6 & 23.7 $\pm$ 2.9 &  $<$10.1 & 7.6 $\pm$ 1.6 & 5.3 $\pm$ 1.3 \\
$^{12}$CO J = 9--8 & 1036.912 &  &  $<$13.6 &  $<$11.7 & 14.6 $\pm$ 2.3 &  $<$12.1 &  $<$15.1 &  $<$8.1 \\
$^{12}$CO J = 10--9 & 1151.985 &  &  $<$13.3 & 13.1 $\pm$ 2.5 & 15.7 $\pm$ 2.4 &  $<$12.9 &  $<$12.2 &  $<$9.8 \\
$^{12}$CO J = 11--10 & 1267.015 &  &  $<$11.6 & 9.7 $\pm$ 2.9 & 13.9 $\pm$ 2.4 &  $<$13.6 &  $<$17.8 &  $<$9.0 \\
$^{12}$CO J = 12--11 & 1381.995 &  &  $<$12.1 &  $<$8.7 &  $<$14.2 &  $<$13.0 &  $<$15.0 &  $<$9.0 \\
$^{12}$CO J = 13--12 & 1496.923 &  &  $<$16.7 &  $<$11.5 &  $<$14.4 &  $<$15.6 &  $<$19.7 &  $<$11.4 \\
 \\
p-H$_2$O 2$_{11}$--2$_{02}$ & 752.033 &  &  $<$5.8 & 8.5 $\pm$ 1.8 &  $<$7.5 &  $<$5.3 &  $<$5.3 &  $<$3.1 \\
p-H$_2$O 2$_{02}$--1$_{11}$ & 987.927 &  &  $<$18.2 &  $<$11.0 &  $<$14.4 &  $<$15.3 &  $<$20.7 &  $<$10.9 \\
o-H$_2$O 3$_{12}$--3$_{03}$ & 1097.365 &  &  $<$11.6 &  $<$9.6 &  $<$12.9 &  $<$11.0 &  $<$12.2 &  $<$8.1 \\
p-H$_2$O 1$_{11}$--0$_{00}$ & 1113.343 &  &  $<$13.0 &  $<$11.4 &  $<$13.7 &  $<$10.6 &  $<$12.2 &  $<$8.4 \\
o-H$_2$O 3$_{21}$--3$_{12}$ & 1162.912 &  &  $<$14.1 & 10.8 $\pm$ 2.4 &  $<$11.2 &  $<$12.0 &  $<$12.3 &  $<$10.2 \\
p-H$_2$O 2$_{20}$--2$_{11}$ & 1228.789 &  &  $<$11.9 &  $<$12.9 & 12.0 $\pm$ 3.2 &  $<$11.2 &  $<$16.4 &  $<$9.5 \\
\\ \hline \\
$[$\ion{C}{1}$]$ $^{3}$P$_{1}$-$^{3}$P$_{0}$ & 492.161 &  &  $<$12.2 &  $<$16.0 & 22.7 $\pm$ 4.3 &  $<$13.2 &  $<$11.3 &  $<$6.9 \\
$[$\ion{C}{1}$]$ $^{3}$P$_{2}$-$^{3}$P$_{1}$ & 809.342 &  & 7.7 $\pm$ 1.5 & 15.8 $\pm$ 2.1 & 46.6 $\pm$ 1.9 & 10.0 $\pm$ 1.5 & 7.8 $\pm$ 1.3 & 8.4 $\pm$ 1.0 \\
$[$\ion{N}{2}$]$ $^{3}$P$_{1}$-$^{3}$P$_{0}$ & 1461.128 &  & 50.5 $\pm$ 3.4 & 37.9 $\pm$ 4.4 & 38.6 $\pm$ 4.3 & 88.1 $\pm$ 3.7 & 25.0 $\pm$ 4.3 & 21.2 $\pm$ 3.1
\enddata
\tablecomments{Measured line fluxes and 1$\sigma$ statistical uncertainties. For the non-detections we state the 3$\sigma$ upper limits. $^{(a)}$ Due to the higher redshift of UGC~05101 the \CO4\ transition is outside the \FTS\ range. }
\end{deluxetable*}
\begin{deluxetable*}{lcccccccc}
\addtocounter{table}{-1}
\tabletypesize{\small}
\tablewidth{0pt}
\tablecaption{SPIRE\slash FTS Line Fluxes -- Continued}
\tablehead{ 
\colhead{Transition}& \colhead{\rm $\nu_{\rm rest}$} & \colhead{} & \multicolumn{5}{c}{\rm Fluxes (10$^{-15}$\,erg\,cm$^{-2}$\,s$^{-1}$)} \\
 \cline{4-8} \\
  &   (GHz)     & & NGC~4388 & IC~3639 & NGC~7130 & NGC~7172 & NGC~7582
}
\startdata
$^{12}$CO J = 4--3 & 461.041 &  & 23.9 $\pm$ 2.9 & 12.1 $\pm$ 1.9 & 35.3 $\pm$ 3.9 &  $<$53.9 & 69.3 $\pm$ 7.2 \\
$^{12}$CO J = 5--4 & 576.268 &  & 19.1 $\pm$ 2.4 &  $<$10.3 & 36.6 $\pm$ 3.9 &  $<$43.7 & 78.0 $\pm$ 8.0 \\
$^{12}$CO J = 6--5 & 691.473 &  & 15.4 $\pm$ 1.9 &  $<$5.4 & 27.7 $\pm$ 3.0 & 21.2 $\pm$ 2.8 & 82.3 $\pm$ 8.3 \\
$^{12}$CO J = 7--6 & 806.652 &  & 13.1 $\pm$ 1.7 &  $<$4.8 & 27.8 $\pm$ 2.9 &  $<$19.0 & 65.6 $\pm$ 6.7 \\
$^{12}$CO J = 8--7 & 921.800 &  & 14.1 $\pm$ 1.8 &  $<$7.3 & 22.2 $\pm$ 2.6 &  $<$27.9 & 63.1 $\pm$ 6.5 \\
$^{12}$CO J = 9--8 & 1036.912 &  & 9.7 $\pm$ 1.6 &  $<$9.6 & 25.2 $\pm$ 2.8 &  $<$9.3 & 54.6 $\pm$ 5.7 \\
$^{12}$CO J = 10--9 & 1151.985 &  &  $<$8.1 &  $<$9.2 & 18.8 $\pm$ 2.4 &  $<$10.4 & 35.1 $\pm$ 3.8 \\
$^{12}$CO J = 11--10 & 1267.015 &  &  $<$10.8 &  $<$11.2 & 12.9 $\pm$ 1.8 &  $<$12.3 & 26.4 $\pm$ 3.0 \\
$^{12}$CO J = 12--11 & 1381.995 &  &  $<$8.0 &  $<$8.6 & 10.3 $\pm$ 1.6 &  $<$9.3 & 10.9 $\pm$ 1.9 \\
$^{12}$CO J = 13--12 & 1496.923 &  &  $<$12.2 &  $<$11.4 &  $<$12.4 &  $<$11.9 &  $<$22.4 \\
 \\
p-H$_2$O 2$_{11}$--2$_{02}$ & 752.033 &  &  $<$5.8 &  $<$4.6 & 6.6 $\pm$ 0.9 &  $<$18.1 &  $<$9.2 \\
p-H$_2$O 2$_{02}$--1$_{11}$ & 987.927 &  &  $<$11.0 &  $<$11.7 & 18.6 $\pm$ 2.0 &  $<$12.1 &  $<$16.9 \\
o-H$_2$O 3$_{12}$--3$_{03}$ & 1097.365 &  &  $<$8.1 &  $<$9.6 & 12.2 $\pm$ 1.5 &  $<$9.0 &  $<$14.0 \\
p-H$_2$O 1$_{11}$--0$_{00}$ & 1113.343 &  &  $<$8.3 &  $<$8.8 &  $<$9.5 &  $<$9.6 &  $<$11.8 \\
o-H$_2$O 3$_{21}$--3$_{12}$ & 1162.912 &  &  $<$16.7 &  $<$8.2 &  $<$12.8 &  $<$18.7 & 13.9 $\pm$ 2.1 \\
p-H$_2$O 2$_{20}$--2$_{11}$ & 1228.789 &  &  $<$11.0 &  $<$10.0 & 12.8 $\pm$ 1.1 &  $<$9.9 &  $<$13.1 \\
\\ \hline \\
$[$\ion{C}{1}$]$ $^{3}$P$_{1}$-$^{3}$P$_{0}$ & 492.161 &  & 14.5 $\pm$ 2.0 &  $<$11.7 & 17.9 $\pm$ 2.2 & 38.2 $\pm$ 5.2 & 34.4 $\pm$ 2.8 \\
$[$\ion{C}{1}$]$ $^{3}$P$_{2}$-$^{3}$P$_{1}$ & 809.342 &  & 26.7 $\pm$ 0.7 & 5.8 $\pm$ 0.7 & 28.2 $\pm$ 1.0 & 45.0 $\pm$ 4.0 & 75.9 $\pm$ 2.0 \\
$[$\ion{N}{2}$]$ $^{3}$P$_{1}$-$^{3}$P$_{0}$ & 1461.128 &  & 47.4 $\pm$ 1.9 & 45.5 $\pm$ 2.1 & 123.0 $\pm$ 1.8 & 120.1 $\pm$ 4.1 & 191.8 $\pm$ 5.4
\enddata
\end{deluxetable*}

\begin{figure*}[!ht]
\centering
\includegraphics[width=1.0\columnwidth]{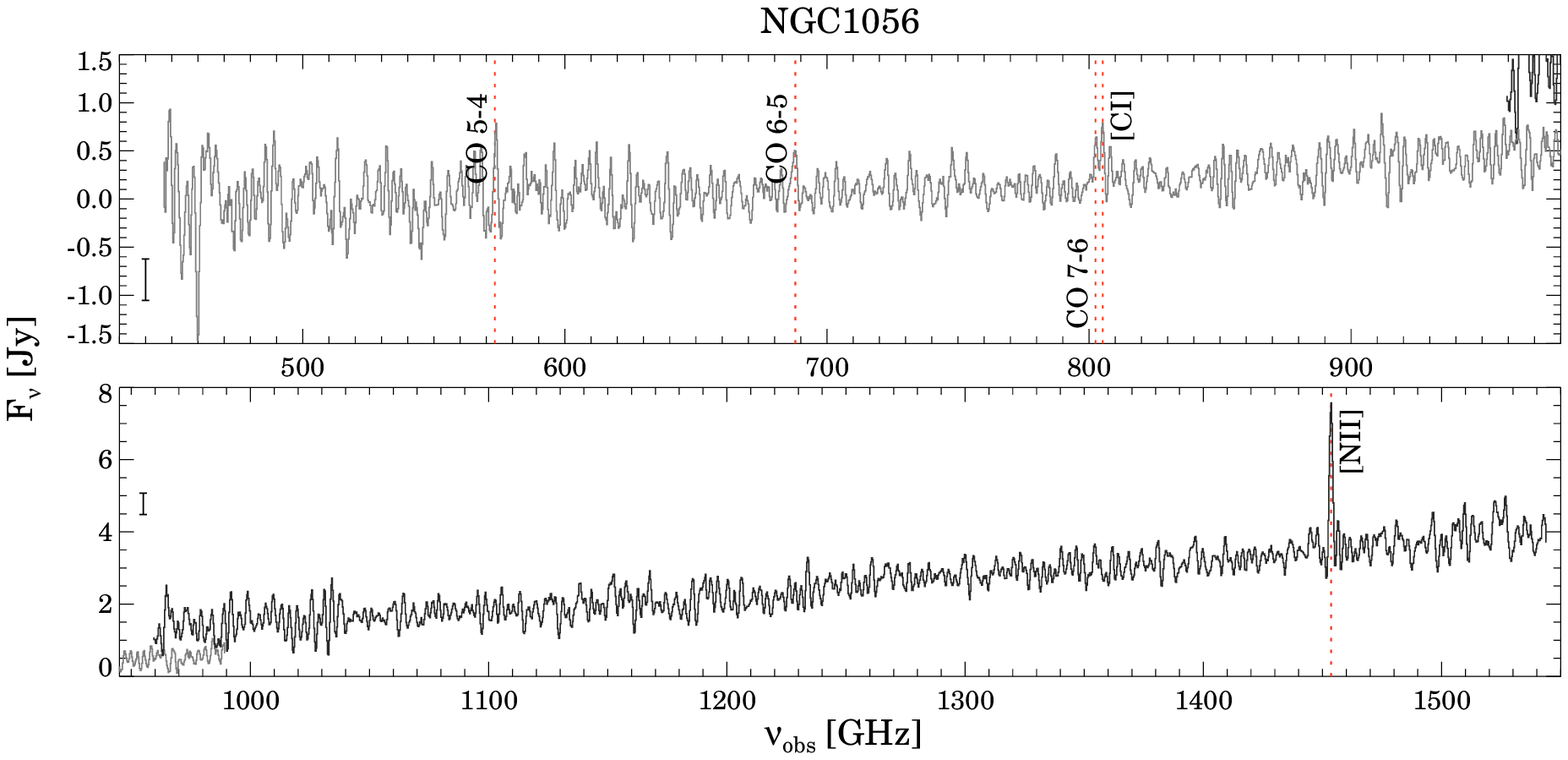}
\includegraphics[width=1.0\columnwidth]{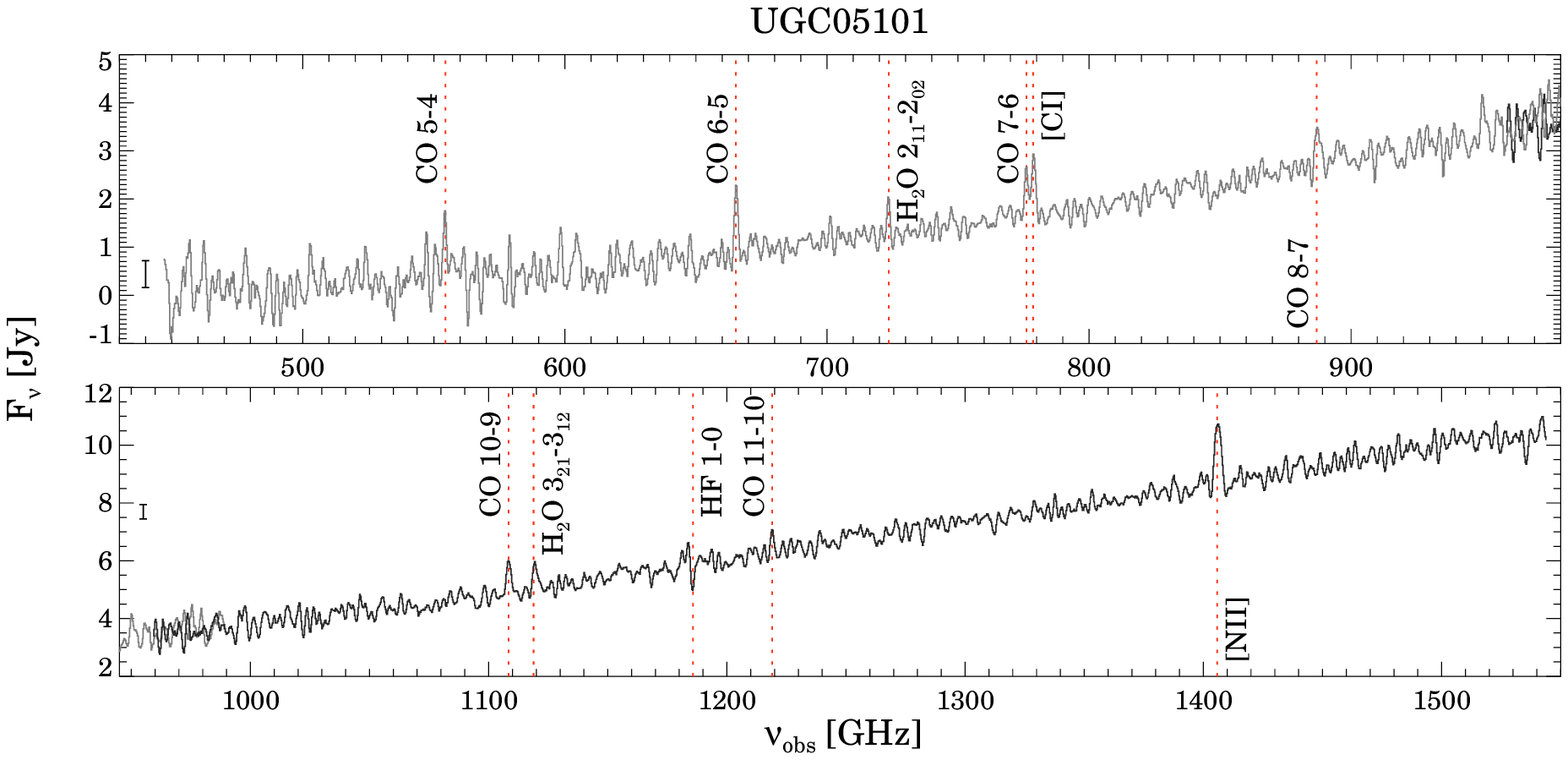}
\includegraphics[width=1.0\columnwidth]{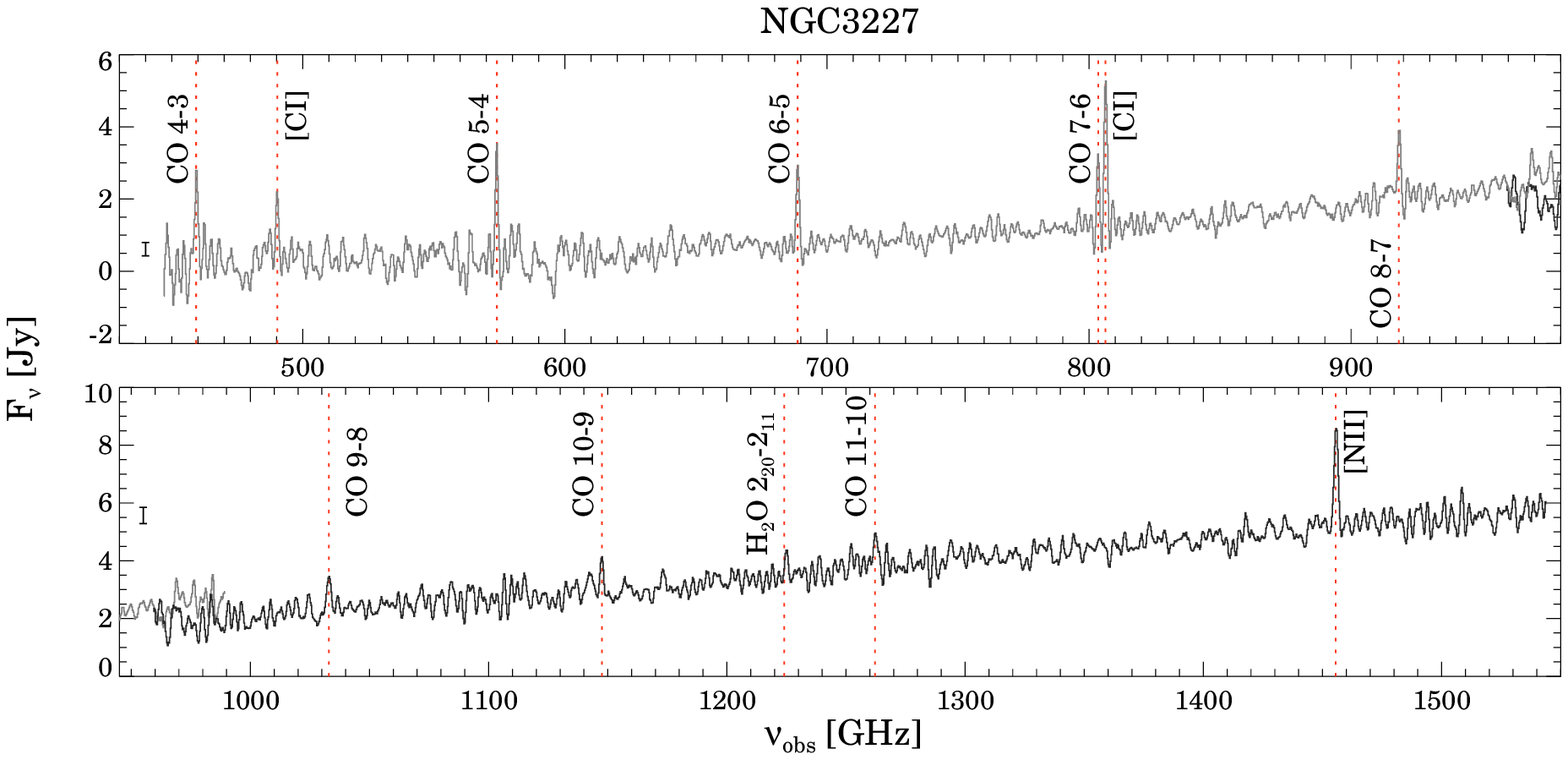}
\includegraphics[width=1.0\columnwidth]{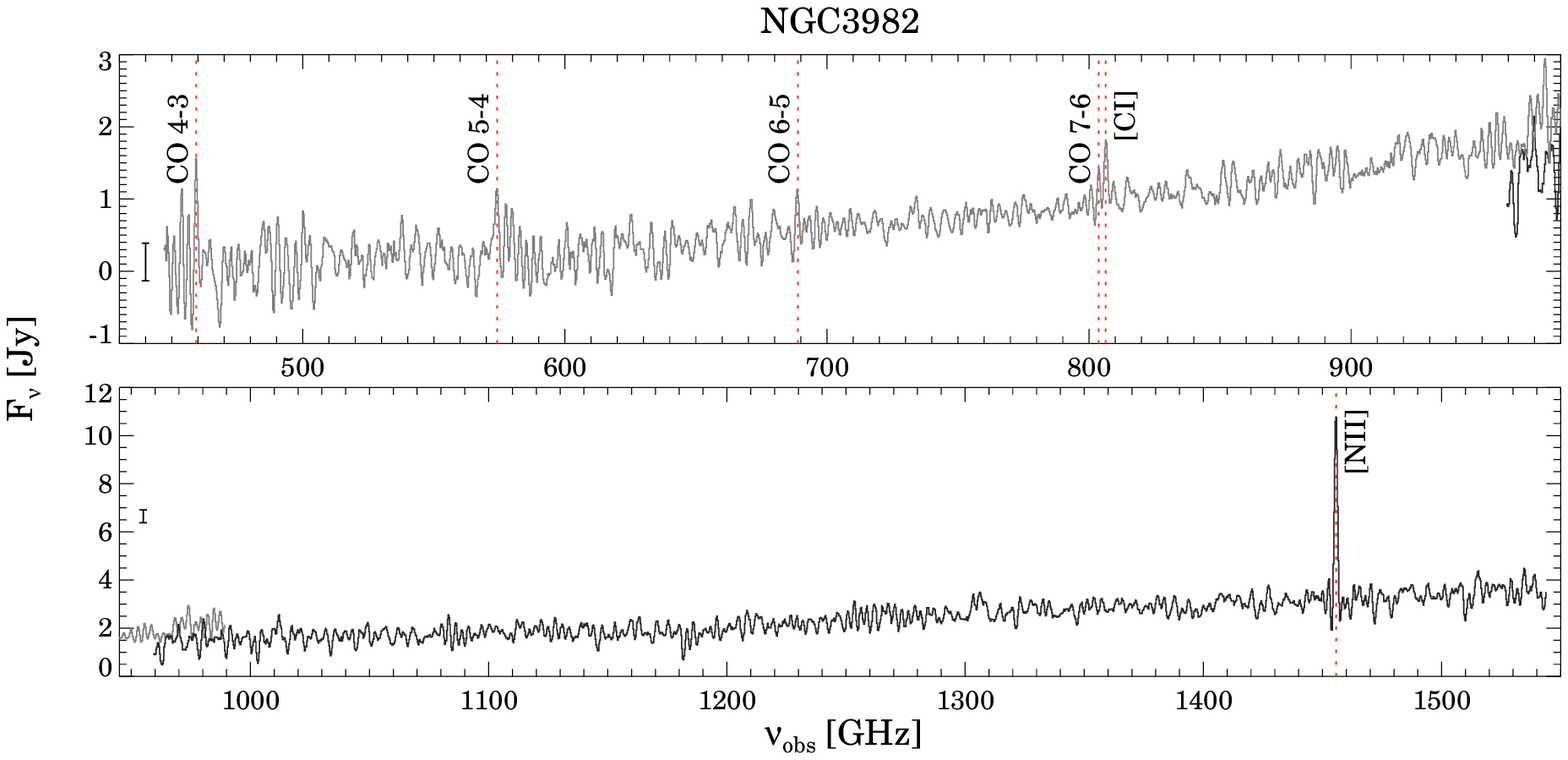}
\includegraphics[width=1.0\columnwidth]{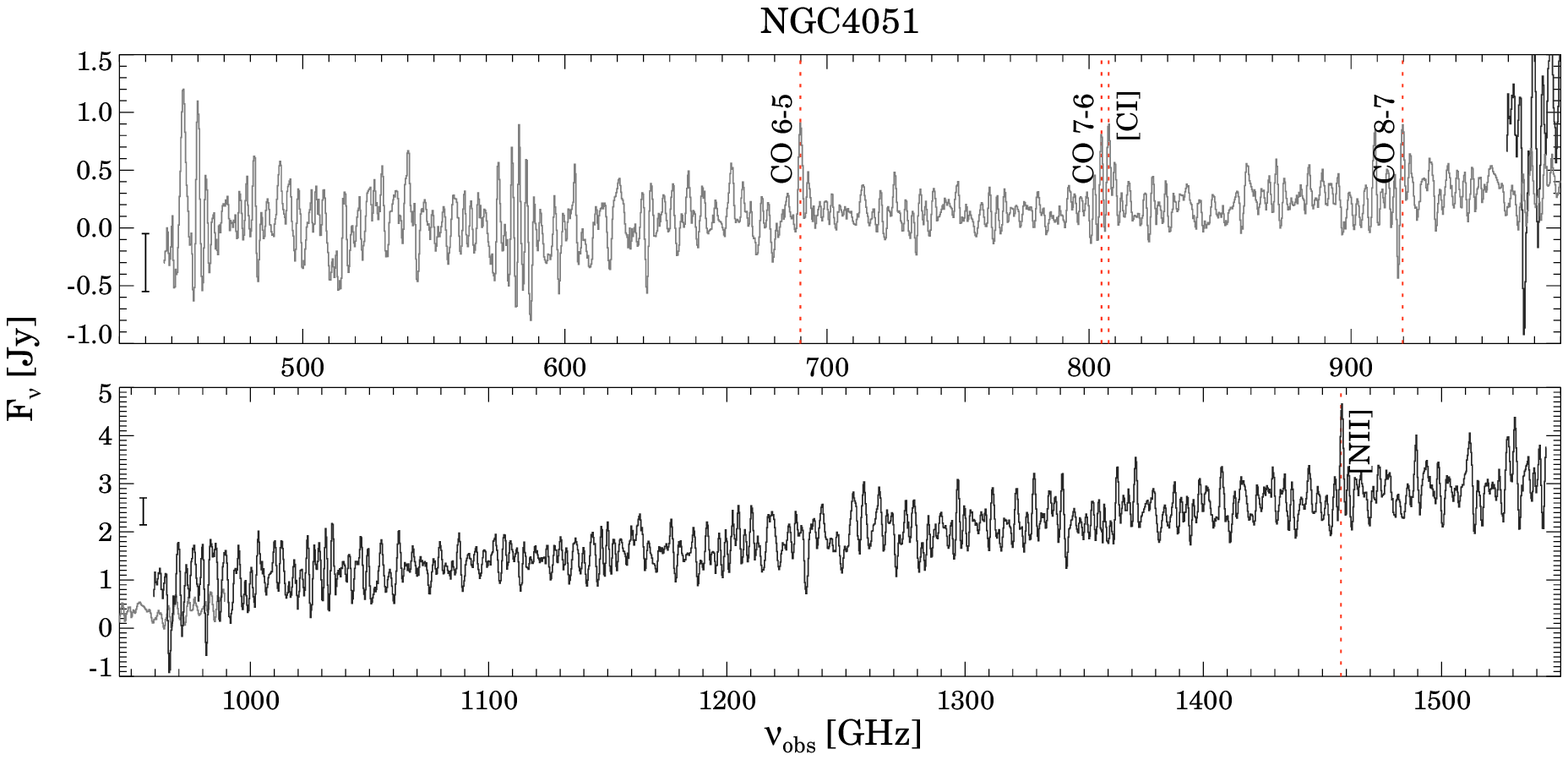}
\includegraphics[width=1.0\columnwidth]{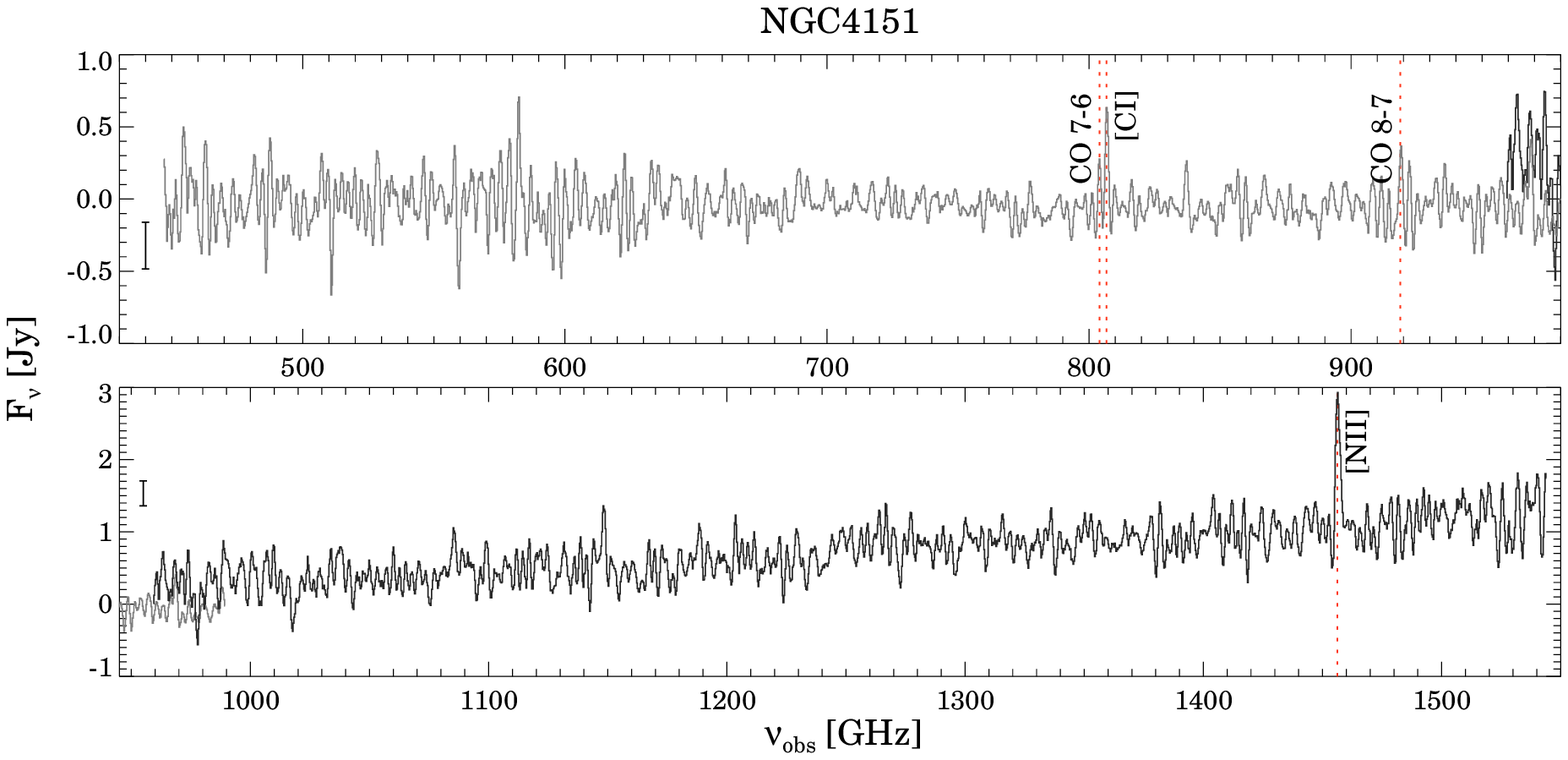}
\caption{Observed \FTS\ spectra of our sample. The black and gray lines are the SSW and SLW spectra, respectively. Notice the overlap between the two spectra in the 960 and 990\,GHz spectral range. The dashed red lines mark the position of the detected lines. The error bars indicate the median 1$\sigma$ uncertainty of each spectrum. \label{fig:spectra}}
\end{figure*}

\begin{figure*}[!ht]
\centering
\addtocounter{figure}{-1}
\includegraphics[width=1.0\columnwidth]{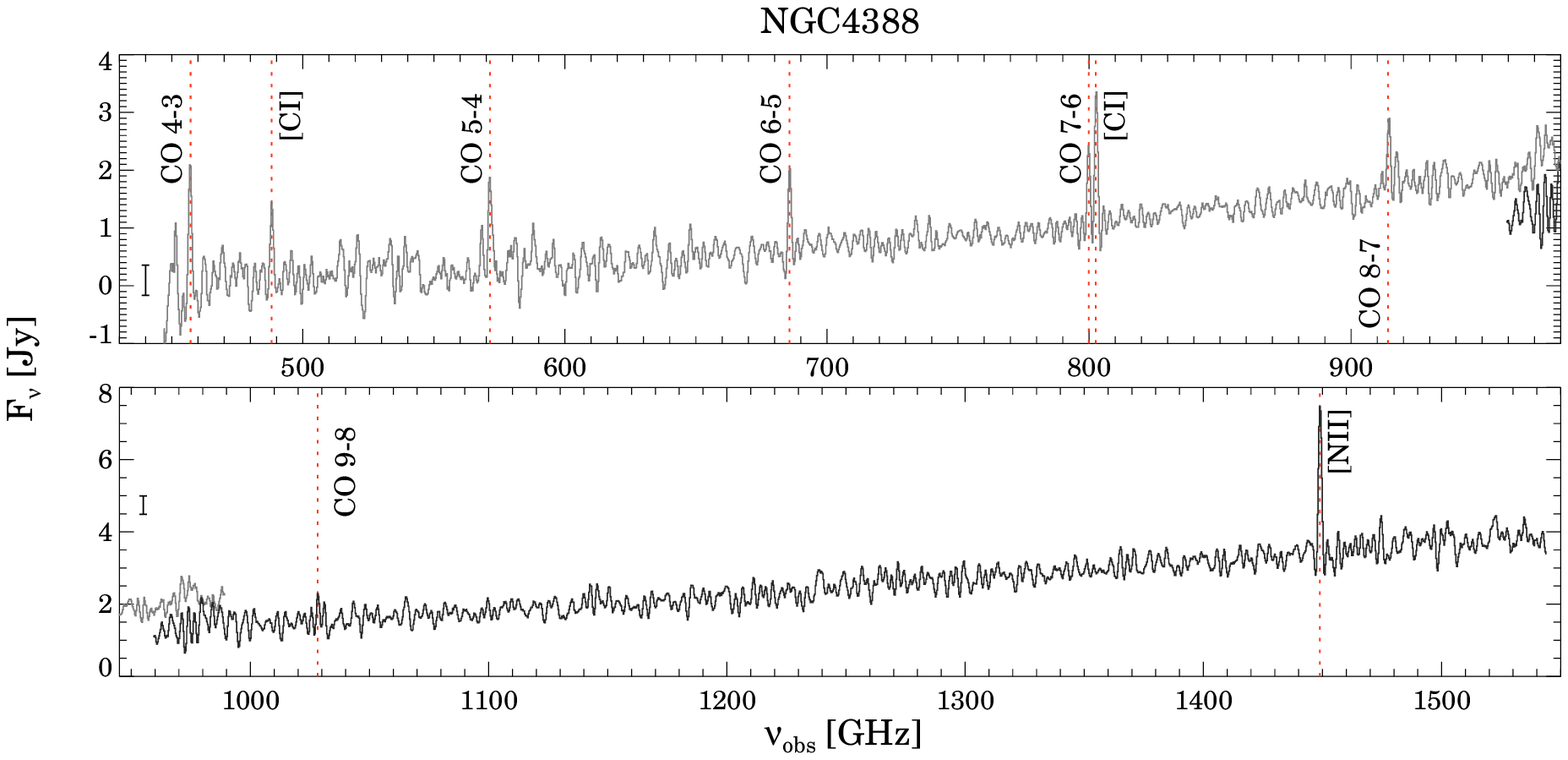}
\includegraphics[width=1.0\columnwidth]{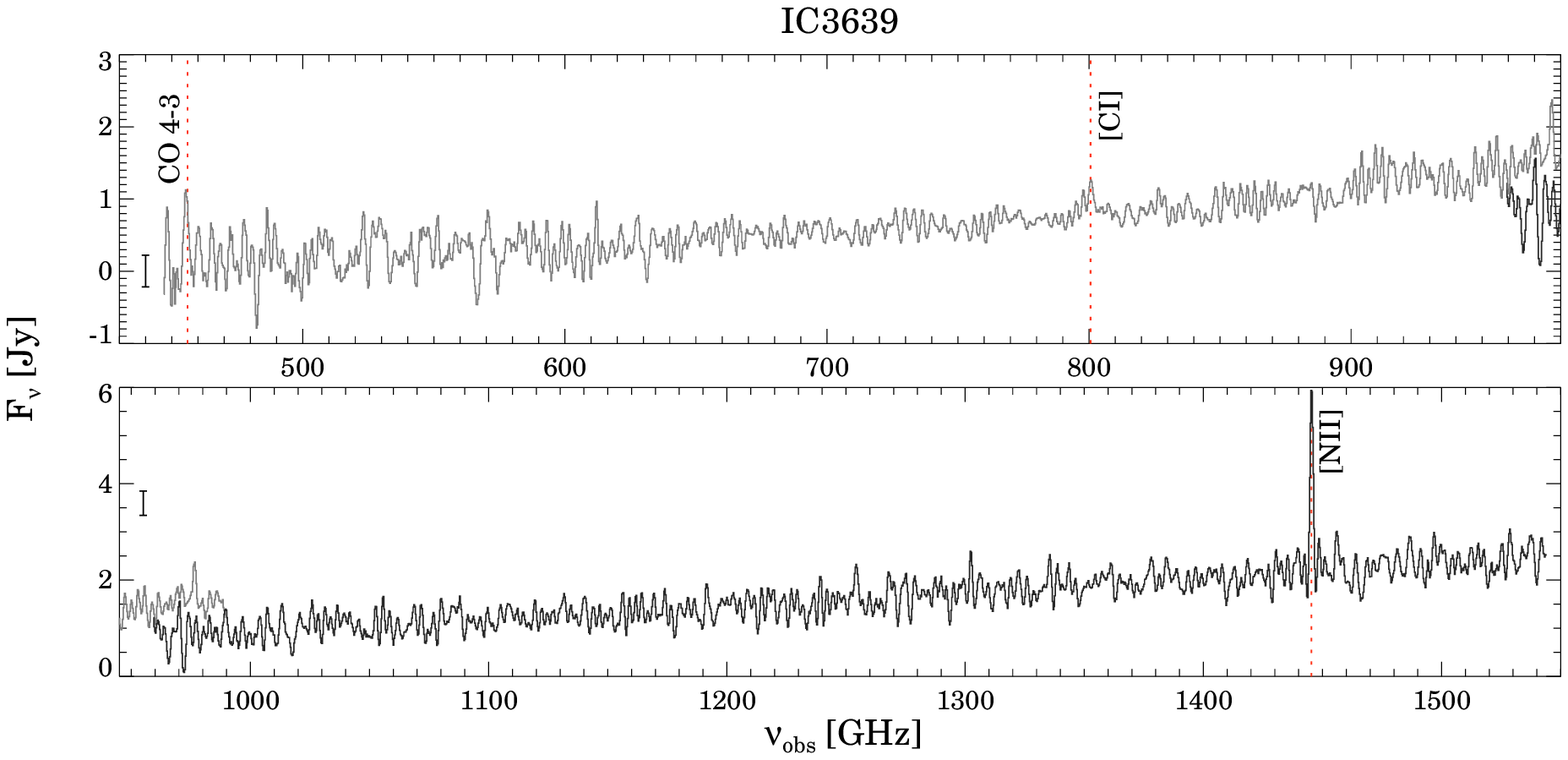}
\includegraphics[width=1.0\columnwidth]{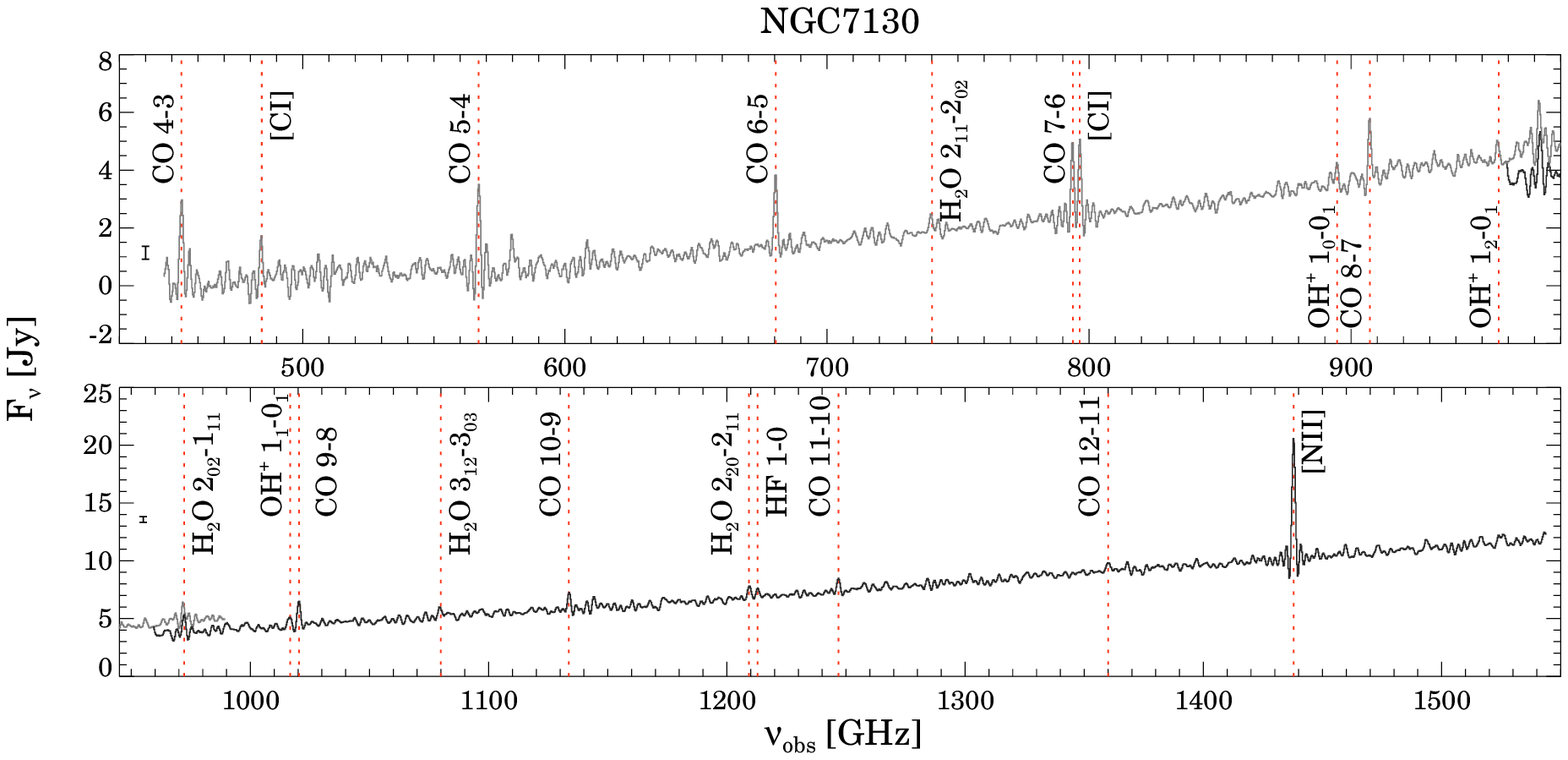}
\includegraphics[width=1.0\columnwidth]{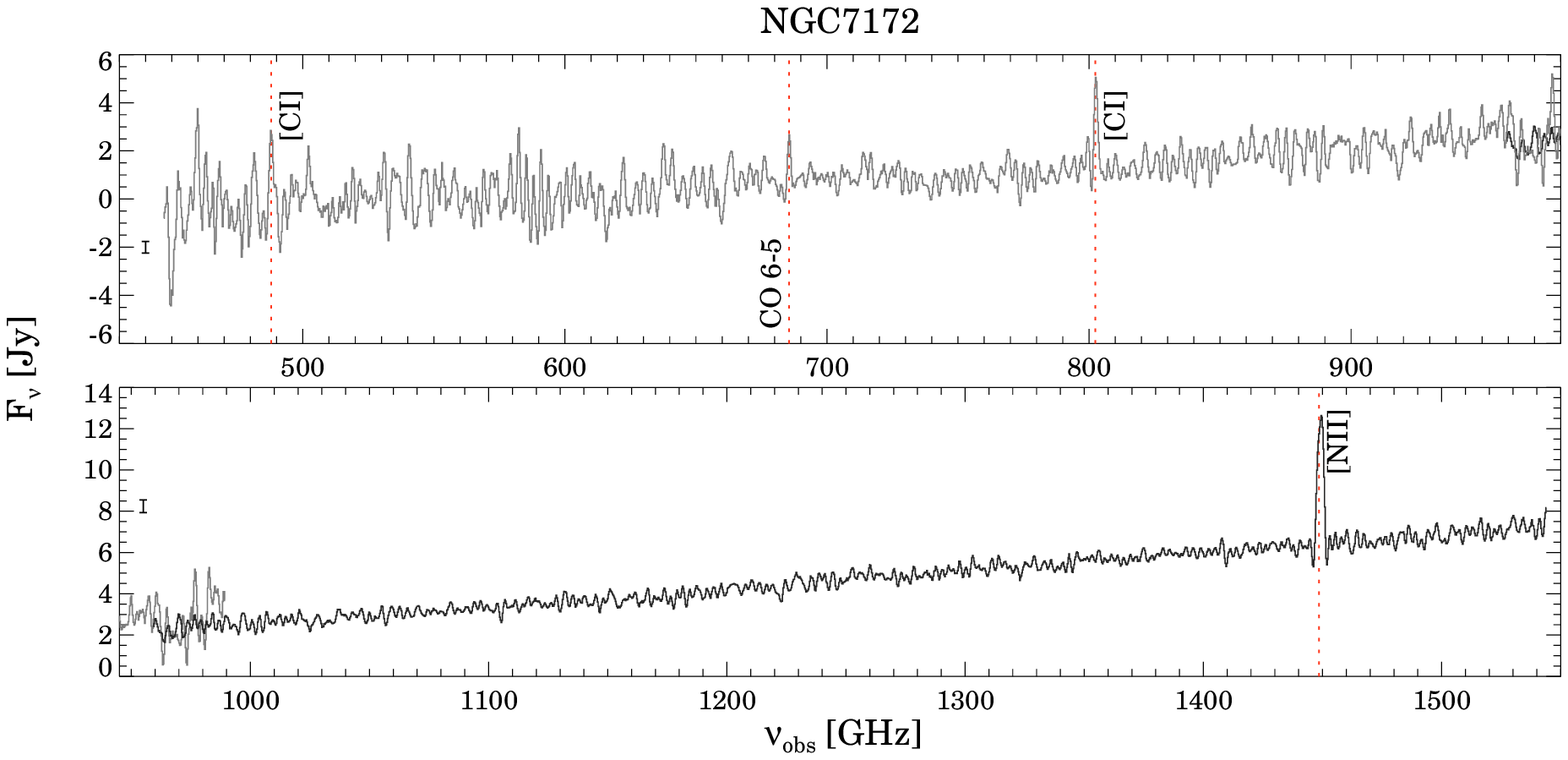}
\includegraphics[width=1.0\columnwidth]{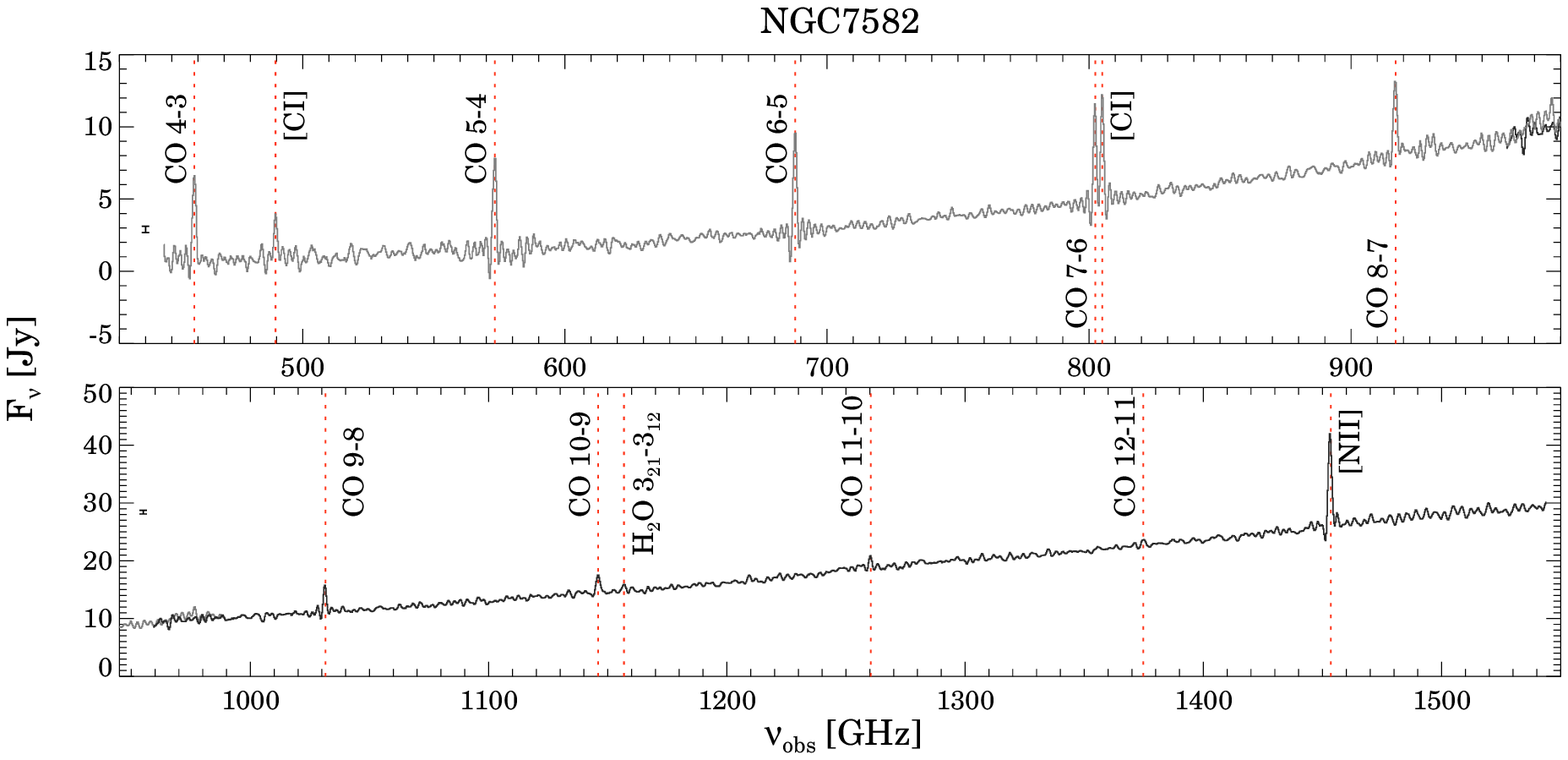}
\hspace{1.0\columnwidth}
\caption{(Continued)}
\end{figure*}

\begin{deluxetable}{lcccccccc}
\tabletypesize{\small}
\tablewidth{0pt}
\tablecaption{SPIRE\slash FTS HF and OH$^+$ Fluxes\label{tab:lines_uncommon}}
\tablehead{ 
\colhead{Galaxy} & \colhead{Transition}& \colhead{\rm $\nu_{\rm rest}$} & \colhead{Flux} \\ 
& & (GHz) & (10$^{-15}$\,erg\,cm$^{-2}$\,s$^{-1}$)}
\startdata
UGC~05101 & HF $J = 1-0$ & 1232.476 & $-$10.4 $\pm$ 2.3\\
\\
NGC~7130 & HF $J = 1-0$ & 1232.476 & 8.6 $\pm$ 1.1 \\
 & OH$^+$ $1_{0}-0_{1}$ & 909.159 & 7.6 $\pm$ 1.2 \\
 & OH$^+$ $1_{2}-0_{1}$ & 971.805 & 9.6 $\pm$ 2.4 \\
 & OH$^+$ $1_{1}-0_{1}$ & 1033.118 & 11.2 $\pm$ 1.5
\enddata
\tablecomments{Fluxes and 1$\sigma$ statistical uncertainties for the HF and OH$^{+}$ detections.}
\end{deluxetable}

\subsection{SPIRE Photometry}\label{ss:photometry}

We obtained \Herschel\ SPIRE imaging observations for seven of our galaxies through a guaranteed time project (PI: L. Spinoglio). The rest of the galaxies in our sample were observed as part of the Very Nearby Galaxies Survey (PI: C.~D. Wilson) and the \Herschel\ Reference Survey (PI: S. Eales) programs and the data were available in the \Herschel\ archive. The SPIRE photometer consists of three bolometer arrays that observe in three spectral bands centered at 250, 350, and 500\,\micron.
Depending on the angular size of the galaxies they were observed in the small or in the large scan-map mode in the three bands.
At least two scans in perpendicular directions were performed for each galaxy.

First we processed the raw data using HIPE version 9 to create the flux calibrated timelines for each bolometer. The standard HIPE pipeline corrects for instrumental effects and attaches the pointing information to the timelines. Then the timelines were combined to create the maps using Scanamorphos version 18 \citep{Roussel2012}. Scanamorphos detects and masks glitches in the data, performs the baseline subtraction, and projects the timelines in a spatial grid.

To measure the nuclear fluxes of the galaxies, first we converted the maps produced by Scanamorphos in Jy\,beam$^{-1}$ to Jy\,arcsec$^{-1}$. We adopted the beam areas given by the SPIRE Observer's Manual (423, 751, and 1587\,arcsec$^2$ for the 250, 350, and 500\,\micron\ bands, respectively). Then we performed aperture photometry using an aperture of diameter 18\arcsec\ in the 250\,\micron\ image and 30\arcsec\ in the 350 and 500\,\micron\ maps and we multiplied the measured fluxes by an aperture correction.
The values of these corrections are 2.33, 1.75, and 3.02 for the 250, 350, and 500\,\micron\ fluxes, respectively. These aperture correction factors were calculated from a map of Neptune processed following the same reduction steps described above. We did not apply any color correction to the fluxes because this correction is expected to be small, $<$5\,\%, for black-body emission (SPIRE Observer's Manual). The measured fluxes are presented in Table \ref{tab:photometry}.

For most of these galaxies, the nuclear region is only marginally resolved at the angular resolution of SPIRE (18\arcsec, 25\arcsec, and 37\arcsec\ for the 250, 350, and 500\,\micron\ bands, respectively). However, for NGC~4051 and NGC~4151 the nuclear emission is more extended, so the reported nuclear flux for these two galaxies might be overestimated due to the aperture correction applied. 
For the two  more distant galaxies in our sample, UGC~05101 and NGC~7130, the fluxes in Table \ref{tab:photometry} correspond to the galaxy integrated sub-mm emission, since they are point sources in the SPIRE images. For the rest, the fluxes should be considered a lower limit to their total sub-mm emission.

\begin{deluxetable}{lccc}[t]
\tablewidth{0pt}
\tablecaption{SPIRE Photometry\label{tab:photometry}}
\tablehead{ \colhead{Galaxy} & \colhead{F$_{\nu}$(250\,\micron)}  & \colhead{F$_{\nu}$(350\,\micron)} & \colhead{F$_{\nu}$(500\,\micron)} \\
& (Jy) & (Jy) & (Jy)}
\startdata
NGC~1056 & 2.15 $\pm$ 0.03 & 1.09 $\pm$ 0.02 & 0.439 $\pm$ 0.012 \\
UGC~05101 & 5.83 $\pm$ 0.11 & 2.35 $\pm$ 0.04 & 0.785 $\pm$ 0.019 \\
NGC~3227 & 3.14 $\pm$ 0.02 & 1.55 $\pm$ 0.01 & 0.611 $\pm$ 0.005 \\
NGC~3982 & 1.89 $\pm$ 0.01 & 1.24 $\pm$ 0.01 & 0.566 $\pm$ 0.009 \\
NGC~4051 & 1.49 $\pm$ 0.02 & 0.780 $\pm$ 0.011 & 0.343 $\pm$ 0.011 \\
NGC~4151 & 0.592 $\pm$ 0.007 & 0.305 $\pm$ 0.005 & 0.133 $\pm$ 0.008 \\
NGC~4388 & 2.39 $\pm$ 0.07 & 1.33 $\pm$ 0.03 & 0.614 $\pm$ 0.019 \\
IC~3639 & 1.56 $\pm$ 0.02 & 0.845 $\pm$ 0.011 & 0.392 $\pm$ 0.012 \\
NGC~7130 & 6.49 $\pm$ 0.10 & 2.92 $\pm$ 0.04 & 1.04 $\pm$ 0.02 \\
NGC~7172 & 4.14 $\pm$ 0.06 & 1.84 $\pm$ 0.03 & 0.711 $\pm$ 0.015 \\
NGC~7582 & 16.2 $\pm$ 0.3 & 6.47 $\pm$ 0.09 & 2.26 $\pm$ 0.04
\enddata
\tablecomments{Nuclear fluxes measured in the SPIRE 250, 350, and 500\,\micron\ images. The aperture size is 18\arcsec\ for the 250\,\micron\ flux and 30\arcsec\ for the 350 and 500\,\micron\ fluxes. Aperture corrections have been applied (see Section \ref{ss:photometry}).}
\end{deluxetable}

\subsection{Ground-based CO Observations}\label{ss:groundCO}

We performed pointed observations of low-$J$ CO lines in three galaxies, UGC\,05101, NGC\,3227, and NGC\,3982. The observations were conducted from 2011 October 1st to October 3rd with the IRAM 30\,m telescope on Pico Veleta (Spain). We used the heterodyne eight mixer receiver (EMIR) using different frequency setups, depending on the target. For NGC\,3227 and NGC\,3982 we used two frequency setups, one centered at 98.875\,GHz for the 3\,mm band (E0) to encompass the emission of \CO1 and $^{13}$\CO1, and another one that combines the 3\,mm band (E0) and 1\,mm band (E2), centered at 89.375\,GHz and 230.538\,GHz, respectively, to cover the emission of the \CO2 line. For UGC\,05101 the receiver was tuned at 98.875\,GHz and 230.538\,GHz to cover the \CO1 and \CO2 lines. The $^{13}$\CO1 was outside the observed bands for UGC\,05101. The 16~GHz bandwidth of the EMIR receivers were connected to the 195~kHz resolution Fourier transform spectrometers (FTS), providing a channel-width spacing of $\sim$0.6\,\kms\ for the 3\,mm band and $\sim$0.3\,\kms\ for the 1\,mm band, and to the wideband line multiple autocorrelator (WILMA) that provides a resolution of 2\,MHz ($5.2-6.7$\,\kms\ for the 3\,mm band and $\sim$2.6\,\kms\ at 230.538\,GHz). The half power beam width (HPBW) is $21\farcs3$ at 115\,GHz and $10\farcs7$ at 230\,GHz. Observations were performed in wobbler switching mode, with a throw of 120\arcsec. Weather conditions were good, with zenith optical depths between 0.1 and 0.3. System temperatures ranged from 100\,K to 200\,K in the 3\,mm band and from 250\,K to 800\,K in the 1\,mm band. Pointing was checked every two hours and the rms pointing error was 4\arcsec\ at 3\,mm. Focus was checked at the beginning of each run and during sunrise.

The data were reduced using the continuum and line analysis single-dish Software (CLASS) package of the Grenoble image and line data analysis software (GILDAS)\footnote{http://www.iram.fr/IRAMFR/GILDAS}. The observed spectra were converted from antenna temperature units ($T_{\mathrm{a}}^{*}$) to main beam temperatures ($T_{\mathrm{mb}}$) using the relation $T_{\mathrm{mb}}=T_{\mathrm{a}}^{*}\times(F_{\mathrm{eff}}/B_{\mathrm{eff}})$, where $F_{\mathrm{eff}}$ is the forward efficiency of the telescope ($F_{\mathrm{eff}}=0.95$ for the 3\,mm band and $F_{\mathrm{eff}}=0.92$ for the 1\,mm band) and $B_{\mathrm{eff}}$ is the main beam efficiency, ranging from 0.77 to 0.81 at 3\,mm, and $B_{\mathrm{eff}}=0.59$ at 1\,mm. A second order baseline was removed for each individual scan and co-added after inspection. The final spectra were smoothed to a common velocity resolution of 20\,\kms. In this work we present the results of the CO and $^{13}$CO lines. The other molecular species detected within the EMIR bands will be presented and analyzed in a forthcoming paper.

In addition, we analyzed the \CO3 transition for these galaxies observed with the 15\,m James Clerk Maxwell Telescope (JCMT) on Mauna Kea (USA) and the 10\,m Heinrich Hertz Telescope (HHT) on Mt. Graham (USA). These observations are described in detail by \citet{Wilson2008}, \citet{Iono2009}, and \citet{Mao2010}.

\subsubsection{CO Line Profile Fitting}\label{ss:groundCO_fitting}

The observed low-$J$ CO transitions have broad and complex line profiles suggesting that the CO emission is produced in several regions unresolved within the telescope beam. To estimate the CO intensity arising from each region we fitted multiple Gaussians to the observed line profiles. Three Gaussians reproduce well the observed profiles in these galaxies.

\begin{figure}
\centering
\includegraphics[width=0.8\columnwidth]{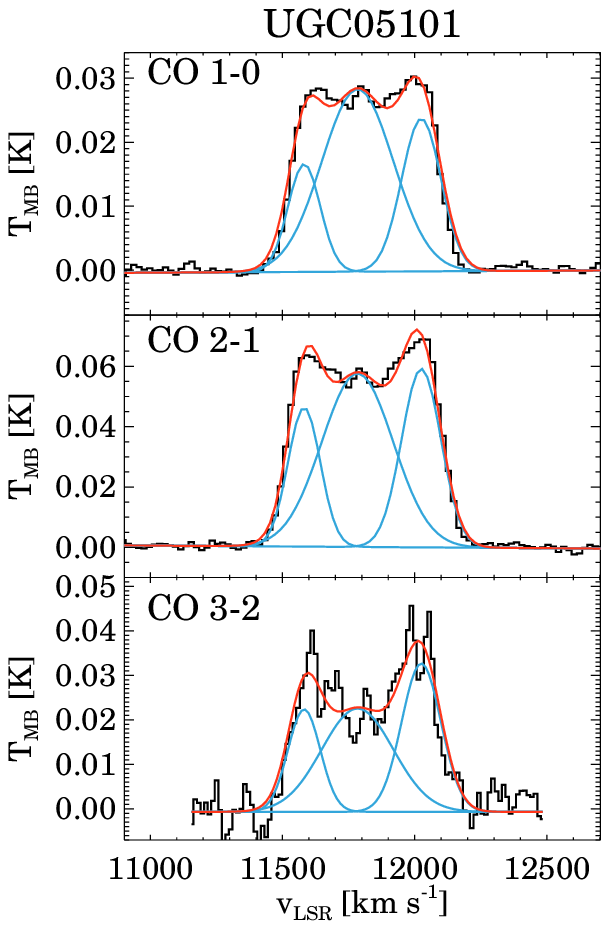}
\caption{Observed line profiles and best-fit models of the \CO1 and \CO2 transitions observed with IRAM 30\,m and the \CO3 transition observed with JCMT 15\,m for UGC~05101. The spectra are smoothed to a velocity resolution of 20\,km\,s$^{-1}$. The narrower spectral coverage of the \CO3 transition is due to the narrower instantaneous bandwidth of the JCMT receivers. \label{fig:iram_fit_ugc5101}}
\end{figure}

\begin{figure}
\centering
\includegraphics[width=0.8\columnwidth]{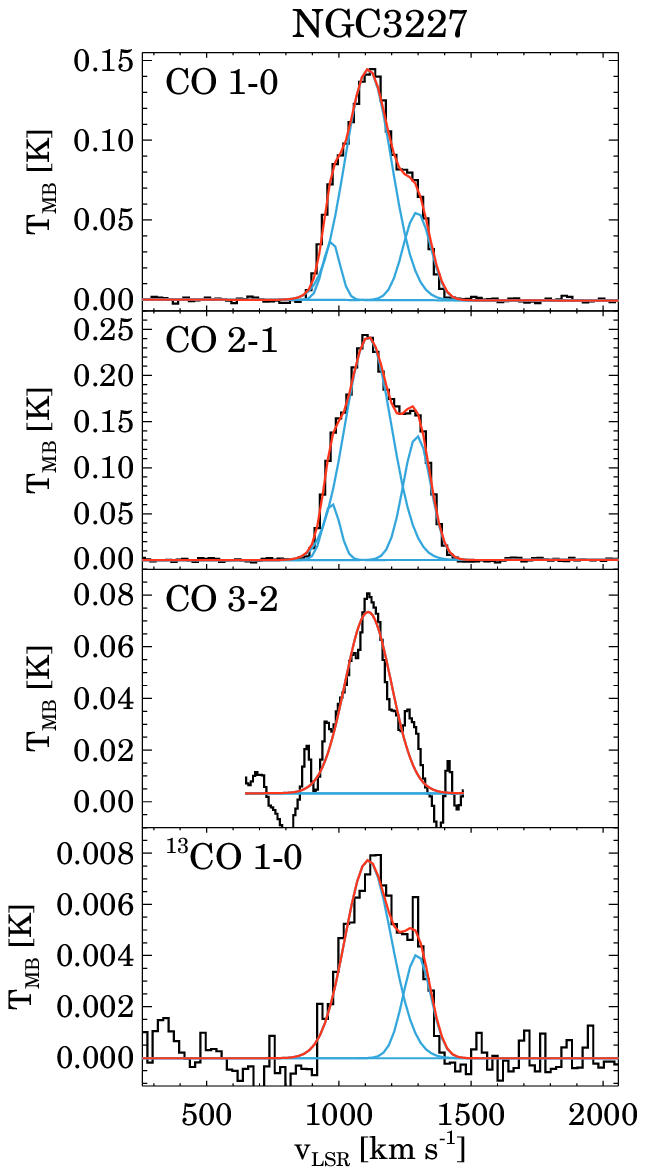}
\caption{Same as Figure \ref{fig:iram_fit_ugc5101}, but for NGC~3227. For this galaxy the \CO3 transition was observed with HHT 10\,m. The narrower spectral coverage of the \CO3 transition is due to the narrower instantaneous bandwidth of the HHT receivers. \label{fig:iram_fit_ngc3227}}
\end{figure}

\begin{figure}
\centering
\includegraphics[width=0.8\columnwidth]{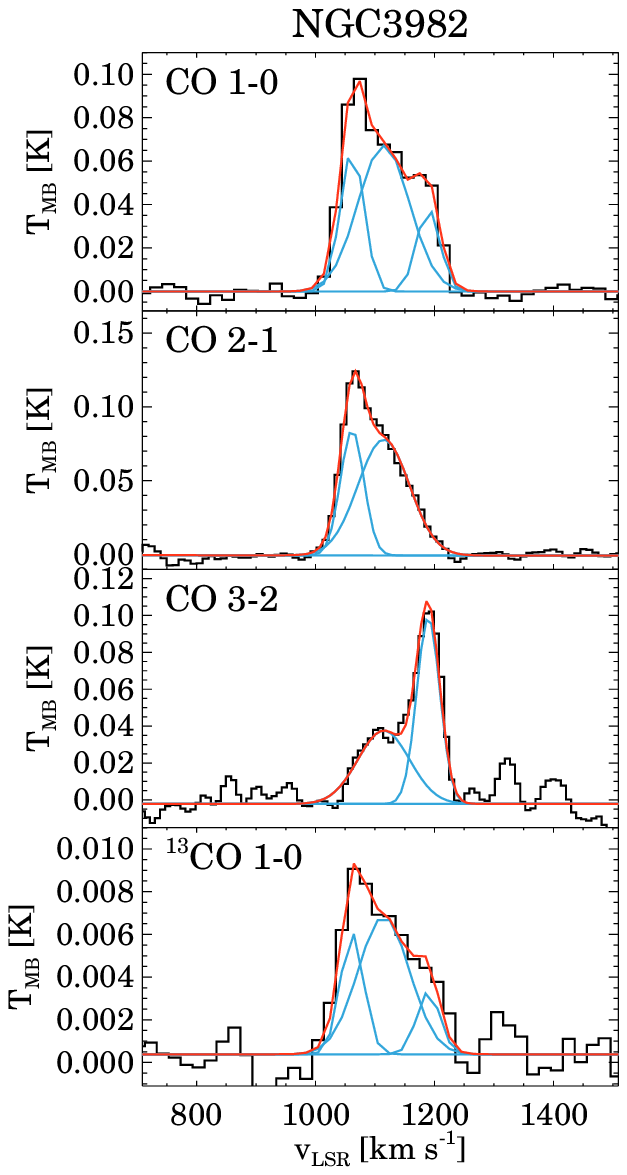}
\caption{Same as Figure \ref{fig:iram_fit_ugc5101}, but for NGC~3982. For this galaxy the \CO3 transition was observed with HHT 10\,m. \label{fig:iram_fit_ngc3982}}
\end{figure}

First, we derived the FWHM and relative velocity of each kinematical component by fitting simultaneously the high signal-to-noise (S\slash N) ratio \CO1 and \CO2 line profiles. We assumed that the FWHM and relative velocity of each kinematical component are equal for both transitions, and only the peak intensities were allowed to vary.
Then, to model the lower S\slash N ratio \CO3 and $^{13}$\CO1 lines we used the FWHM and velocity values derived before and fitted the peak intensities of each component.
The best-fit models are shown in Figures \ref{fig:iram_fit_ugc5101}, \ref{fig:iram_fit_ngc3227}, and \ref{fig:iram_fit_ngc3982}. In Table \ref{tab:lines_iram} we show the results of the Gaussian fits.

\begin{deluxetable*}{lcccccccccc}[ht]
\tabletypesize{\small}
\tablewidth{0pt}
\tablecaption{CO Line Profile Fitting Results\label{tab:lines_iram}}
\tablehead{
\colhead{Galaxy} & ${v_{\rm LSR}}^a$ & FWHM & \colhead{\CO1}  & \colhead{\CO2} & \colhead{\CO3} & \colhead{$^{13}$\CO1} \\
 & (km\,s$^{-1}$) & (km\,s$^{-1}$) & (K\,km\,s$^{-1}$) & (K\,km\,s$^{-1}$) & (K\,km\,s$^{-1}$) & (K\,km\,s$^{-1}$)
}
\startdata

UGC~05101 & 11784 & 318 $\pm$ 32 & 9.64 $\pm$ 0.38 & 19.46 $\pm$ 0.64 & 7.9 $\pm$ 1.6 & \nodata \\
 & 11581 & 144 $\pm$ 8 & 2.58 $\pm$ 0.17 & 7.07 $\pm$ 0.29 & 3.53 $\pm$ 0.72 & \nodata \\
 & 12025 & 175 $\pm$ 5 & 4.45 $\pm$ 0.21 & 11.04 $\pm$ 0.35 & 6.22 $\pm$ 0.87 & \nodata \\

NGC~3227  & 1110 & 206 $\pm$ 8 & 31.83 $\pm$ 0.45 & 53.17 $\pm$ 0.53 & 15.4 $\pm$ 2.0 & 1.70 $\pm$ 0.15 \\
 & 972 & 75 $\pm$ 7 & 2.97 $\pm$ 0.16 & 4.92 $\pm$ 0.19 & $<$ 2.2 & $<$ 0.2 \\
 & 1295 & 122 $\pm$ 4 & 7.21 $\pm$ 0.27 & 17.60 $\pm$ 0.31 & $<$ 3.6 & 0.53 $\pm$ 0.09 \\

NGC~3982 & 1114 & 102 $\pm$ 9 & 7.31 $\pm$ 0.26 & 8.53 $\pm$ 0.33 & 4.3 $\pm$ 1.4 & 0.70 $\pm$ 0.10 \\
& 1061 & 46 $\pm$ 5 & 3.25 $\pm$ 0.12 & 4.20 $\pm$ 0.15 & $<$ 1.8 & 0.28 $\pm$ 0.05 \\
 & 1189 & 45 $\pm$ 8 & 1.87 $\pm$ 0.12 & $<$ 0.5 & 4.94 $\pm$ 0.61 & 0.14 $\pm$ 0.05
\enddata
\tablecomments{$^{(a)}$ The uncertainties of the measured radio local standard of rest (LSR) velocities are $\sim$30\,km\,s$^{-1}$.}
\end{deluxetable*}

\subsubsection{Literature CO Data}\label{ss:literature}

We obtained from the literature the observed intensities of the \CO1, \CO2, and \CO3 transitions for the remaining galaxies (see Table \ref{tab:lines_ground}). For most of them the \CO1 and \CO2 intensities are available. However, the \CO3 transition was observed in just three of them (see also Section \ref{ss:groundCO_fitting}).

For the three galaxies analyzed in Section \ref{ss:groundCO_fitting}, in Table \ref{tab:lines_ground} we list the CO fluxes corresponding to the component with the central recessional velocity. We associate this component with the compact nuclear CO emission, while the other two components are likely due to the rotation of the galaxy disk. The disk emission also contributes to the central component, so the adopted nuclear flux might be slightly overestimated.

\begin{deluxetable*}{lcccccccc}
\tabletypesize{\small}
\tablewidth{0pt}
\tablecaption{CO Line Fluxes from Ground-based Observations\label{tab:lines_ground}}
\tablehead{ 
\colhead{Galaxy} & \colhead{\CO1} & \colhead{Ref.$^a$} & \colhead{\CO2} & \colhead{Ref.$^a$} & \colhead{\CO3} & \colhead{Ref.$^a$} \\ 
& (10$^{-16}$\,erg\,cm$^{-2}$\,s$^{-1}$) & & (10$^{-16}$\,erg\,cm$^{-2}$\,s$^{-1}$) & & (10$^{-16}$\,erg\,cm$^{-2}$\,s$^{-1}$)}
\startdata
NGC~1056 & 2.5 $\pm$ 0.5 & 1,I & 10.4 $\pm$ 2.1 & 1,I & \nodata & \nodata \\
UGC~05101 & 1.8 $\pm$ 0.4 & 2,I & 9.7 $\pm$ 2.0 & 2,I & 22.5 $\pm$ 6.4 & 3,J \\
NGC~3227 & 6.1 $\pm$ 1.2 & 2,I & 26.5 $\pm$ 5.3 & 2,I & 125 $\pm$ 30 & 4,2,H \\
NGC~3982 & 1.4 $\pm$ 0.3 & 2,I & 4.3 $\pm$ 0.9 & 2,I & 34.9 $\pm$ 13.0 & 4,2,H \\
NGC~4051 & 5.6 $\pm$ 1.2 & 5,N & \nodata & \nodata & \nodata & \nodata \\
NGC~4151 & \nodata & \nodata & 3.0 $\pm$ 1.2 & 6,J & \nodata & \nodata \\
NGC~4388 & 1.7 $\pm$ 0.4 & 7,I & 5.5 $\pm$ 1.7 & 7,I & \nodata & \nodata \\
IC~3639 & 2.2 $\pm$ 0.4 & 1,S & 20.9 $\pm$ 4.2 & 1,S & \nodata & \nodata \\
NGC~7130 & 12.4 $\pm$ 2.5 & 1,S & 44.8 $\pm$ 9.0 & 1,S & \nodata & \nodata \\
NGC~7172 & 13.3 $\pm$ 2.7 & 1,S & 26.1 $\pm$ 5.2 & 1,S & \nodata & \nodata \\
NGC~7582 & 18.0 $\pm$ 3.7 & 8,S & 139 $\pm$ 30 & 8,S & \nodata & \nodata 
\enddata
\tablecomments{Fluxes and 1$\sigma$ uncertainties of ground-based observations of the low-$J$ CO lines. The fluxes were converted from their original units (K\,km\,s$^{-1}$) using the following relation: $ F{\rm (erg\,cm^{-2}\,s^{-1})} = I{\rm(K\,km\,s^{-1})}~9.197\times 10^{-13} \nu{\rm (GHz)} \slash (\pi \eta_{\rm MB} (R{\rm (cm)})^2)$, where $\nu$ is the transition frequency, $\eta_{\rm MB}$ the main beam efficiency, and $R$ the radius of the telescope.\\
$^{(a)}$ Reference for the CO line flux and telescope used. (I) IRAM 30\,m; (J) JCMT 15\,m; (H) HHT 10\,m; (N) NRO 45\,m; (S) SEST 15\,m.}

\tablerefs{(1) \citet{Albrecht2007}. (2) This work (Section \ref{ss:groundCO}). (3) \citet{Wilson2008}. (4) \citet{Mao2010}. (5) \citet{Vila-Vilaro1998}. (6) \citet{Rigopoulou1997}. (7) \citet{Pappalardo2012}. (8) \citet{Aalto1995}.}
\end{deluxetable*}

\section{CO Emission Modeling }\label{s:rad_transfer}

In this section we describe the analysis of the CO spectral line energy distributions (SLEDs) of six of our galaxies with more than three CO lines detected in their \FTS\ spectra. First we describe the radiative transfer models used to interpret the CO SLEDs. The models for individual sources are discussed later. Finally we derive the properties of the cold and warm molecular gas traced by the CO lines.

\subsection{Radiative Transfer Models}\label{s:rad_trans}

We used the non local thermodynamic equilibrium (non-LTE) radiative transfer code RADEX \citep{vanderTak2007} to infer the physical conditions in the regions where the mid-$J$ CO emission is produced. 
RADEX uses the escape probability approximation to solve iteratively for the molecular level populations and the intensities of the CO lines.

We produced a grid of models assuming a spherically symmetric homogeneous medium. These models span a wide range in kinetic temperature ($T_{\rm kin} = 50-3000$\,K), molecular hydrogen density ($n_{\rm H_2} = 10-10^5$\,cm$^{-3}$), and CO column density per unit of line width ($N_{\rm CO}\slash \Delta v = 10^{12}-10^{17}$\,cm$^{-2}$\slash km\,s$^{-1}$).
We adopted the collisional rate coefficients between CO and H$_2$ of \citet{Yang2010}. For the background radiation we used the cosmic microwave background (CMB) at 2.73\,K, the inclusion of the local far-IR background does not modify significantly the models (see \citealt{Spinoglio2012}).

The CO lines are spectroscopically unresolved in the FTS spectra of these galaxies, so we used ground-based observations of the low-$J$ CO lines to determine the value of $\Delta v$ (see Section \ref{ss:radex_individual} for details).

The free parameters of the models are the $T_{\rm kin}$, $n_{\rm H_2}$, $N_{\rm CO}$, and the filling factor $\Phi$. The filling factor represents the fraction of the beam area occupied by the source.
When the emission is optically thin ($\tau \ll 1$), in the escape probability approximation, the line ratios only depend on $T_{\rm kin}$ and $n_{\rm H_2}$, whereas the absolute fluxes depend on the $N_{\rm CO} \Phi$ product. Consequently, it is not possible to determine simultaneously $N_{\rm CO}$ and the filling factor, and since the size of the CO emitting region (i.e., the filling factor) in our galaxies is not accurately known, we can only estimate the beam-averaged column density, $\NCOav= N_{\rm CO} \Phi$.

We estimated the values and uncertainties on the model parameters using a Monte Carlo method. From the fluxes and their associated uncertainties listed in Tables \ref{tab:lines} and \ref{tab:lines_ground} we constructed 500 CO SLED simulations for each galaxy that were fitted using a $\chi^2$ minimization algorithm. The parameter values and 1$\sigma$ uncertainties reported in Table \ref{tab:radexfit} are those obtained from the parameter distributions in the Monte Carlo simulations. The best-fit models of the six analyzed galaxies are shown in Figure \ref{fig:radex_model}.

From the derived column densities we calculated the molecular gas masses using the following expression:
\begin{equation}
\label{eq:masa_mol}
M_{\rm mol} = \mu  <\!N_{\rm CO}\!> {x_{\rm CO}}^{-1} \Omega d^2 m_{\rm H_2}, 
\end{equation}
where $\mu=1.4$ is the mean molecular weight per H atom, $<\!N_{\rm CO}\!>$ is the beam-averaged CO column density, $x_{\rm CO}=3\times10^{-4}$ is the assumed CO abundance with respect to H$_2$ in dense molecular clouds \citep{Lacy1994, Hernande2011}, $\Omega$ is the adopted SPIRE SLW beam size (1.66$\times$10$^{-8}$\,sr), $d$ is the distance to the galaxy, and $m_{\rm H_2}$ is the mass of an H$_2$ molecule. In Table \ref{tab:gasprop} we present the calculated molecular gas masses.

\begin{deluxetable*}{lcccccccccc}
\tablewidth{0pt}
\tabletypesize{\small}
\tablecaption{ \label{tab:radexfit} RADEX Models} 
\tablehead{\colhead{Galaxy} & \colhead{$\Delta v^a$} & \colhead{${\log T_{\rm kin}}^b$} & \colhead{${\log n_{\rm H_2}}^c$} & \colhead{${\log \NCOav}^d$} & \colhead{${F_{\rm CO}}^e$} \\
&  (\kms) & (K) & (cm$^{-3}$) & (cm$^{-2}$)  & (10$^{-15}$\,erg\,cm$^{-2}$\,s$^{-1}$)}
\startdata
UGC~05101 & 320 & 2.9 $\pm$ 0.3 & 3.7 $\pm$ 0.4  & 15.2 $\pm$ 0.1  & 102 $\pm$ 11 \\
NGC~3227 & 200 & 2.8 $\pm$ 0.1 & 3.3 $\pm$ 0.1  & 15.8 $\pm$ 0.1   & 190 $\pm$ 10   \\
NGC~3982 & 100 & 2.0 $\pm$ 0.1 & 4.5 $\pm$ 0.4  & 15.2 $\pm$ 0.2   &  39 $\pm$ 4   \\
NGC~4388 & 125 & 2.5 $\pm$ 0.2 & 3.9 $\pm$ 0.2  & 15.4 $\pm$ 0.1   & 104 $\pm$ 12 \\
NGC~7130 & 90 & 2.8 $\pm$ 0.1 & 3.2 $\pm$ 0.1  & 16.1 $\pm$ 0.1   & 240 $\pm$ 10 \\
NGC~7582 & 200 & 2.6 $\pm$ 0.2 & 3.6 $\pm$ 0.7  & 16.4 $\pm$ 0.2   & 540 $\pm$ 30
\enddata
\tablecomments{Best-fit parameters of the RADEX models. $^{(a)}$ Line width assumed for the model. $^{(b)}$ Kinetic temperature. $^{(c)}$ Molecular hydrogen density. $^{(d)}$ Beam-averaged CO column density. The beam FWHM is 30\arcsec. $^{(e)}$ Integrated flux of the CO lines from $J_{\rm up}=4$ to $J_{\rm up}=12$ of the best-fit model.}
\end{deluxetable*}

\subsection{Analysis of Individual Sources}\label{ss:radex_individual}

\subsubsection{UGC~05101}

UGC~05101 is an IR bright galaxy ($\Lir = 10^{11.95}\,\Lsun$) located at 163\,Mpc \citep{SandersRBGS}. It has a peculiar morphology that suggests a recent interaction \citep{Sanders1988}.
Its nuclear activity is classified as LINER from optical spectroscopy \citep{Yuan2010}.
However the detection of high-ionization emission lines in the mid-IR ([\ion{Ne}{5}] and [\ion{O}{4}]; \citealt{Armus07}), the intense Fe\,K$\alpha$ emission line at 6.4\,keV \citep{Imanishi2003b}, and its high hard X-ray luminosity ($L_{\rm 2-10\,keV} = 7.5\times 10^{42}$\,erg\,s$^{-1}$; \citealt{Imanishi2003b}) indicate the presence of a buried AGN with a luminosity comparable to that of a Seyfert galaxy.

In the \FTS\ spectrum of this galaxy we detect five unblended CO lines (see Table \ref{tab:lines}). An emission line is detected at the frequency of the \CO{10} transition, but it is likely blended with the o-H$_2$O\,3$_{12}$--2$_{21}$ 1153\,GHz line since water and CO emission lines have comparable fluxes in this galaxy. With the spectral resolution of \FTS\ it is not possible to deblend the two emission lines; therefore we excluded it from the radiative transfer analysis.

In addition to the \FTS\ data, we used ground-based observations of the low-$J$ CO transitions to constrain the radiative models. Interferometric observations of the \CO{1} line show that the CO emission is produced in a compact region ($<$1\,kpc) plus a fainter extended (2.2\,kpc) rotating disk \citep{Genzel1998,Wilson2008}.

The observed FWHM of the \CO{3} line is 650\,\kms\ \citep{Iono2009}. However, the \FTS\ mid-$J$ CO lines are spectroscopically unresolved, that is, the width of mid-$J$ lines is $<$500\,\kms.
The line profile fitting of the \CO{1}, $J=2-1$, and $J=3-2$ transitions (Section \ref{ss:groundCO_fitting}) shows that the width of the middle component is 320\,\kms, so we assumed that it is also the width of the mid-$J$ CO lines. Consequently, we used the \CO{1}, $J=2-1$, and $J=3-2$ fluxes of this middle component in the CO SLED fit presented in Figure \ref{fig:radex_model}.

\subsubsection{NGC~3227}

NGC~3227 is a nearby (14.4\,Mpc) Seyfert 1.5 galaxy \citep{Ho1997}. It is interacting with the dwarf elliptical galaxy NGC~3226. The hard X-ray luminosity of the AGN is 1.9$\times$10$^{41}$\,erg\,s$^{-1}$ \citep{Gondoin2003}. In addition, recent ($\sim$40\,Myr) intense star-formation is observed in the nucleus. The luminosity produced by these young stars represents about half of the bolometric luminosity of the galaxy \citep{Davies2006}. The \Spitzer\slash IRS mid-IR spectrum of this galaxy shows the [\ion{Ne}{5}] and [\ion{O}{4}] high-ionization emission lines confirming the presence of an AGN \citep{Deo2007, Dasyra2008, Weaver2010}.

Interferometric observations of this galaxy \citep{Schinnerer2000} show that part of the \CO{1} and $J=2-1$ emission is concentrated in a compact region of $\sim$8\arcsec\ of diameter ($\sim$500\,pc). 

In the FTS spectrum of NGC~3227 we detected the CO lines from $J_{\rm up} = 4$ to $J_{\rm up} = 11$. These CO emission lines are not resolved at the \FTS\ spectral resolution; thus we assumed the line width of the middle component of the \CO{1} and $J=2-1$ transitions, 200\,\kms (see Section \ref{ss:groundCO_fitting}).

\subsubsection{NGC~3982}

NGC~3982 is a Seyfert 1.9 \citep{Ho1997} located at 20.6\,Mpc. The hard X-ray luminosity of this galaxy is $<$2.5$\times$10$^{39}$\,erg\,s$^{-1}$ and the Fe\,K$\alpha$ feature at 6.4\,keV is not detected \citep{Guainazzi2005}. This suggests that NGC~3982 hosts a relatively low luminosity AGN. Moreover, the luminosities of the high-ionization mid-IR [\ion{O}{4}] and [\ion{Ne}{5}] emission lines are about 2$\times$10$^{39}$\,erg\,s$^{-1}$ \citep{Tommasin2010}, low compared to those of other Seyfert galaxies (see Figure 2 of \citealt{Pereira2010c}) and confirming that the AGN in NGC~3982 is not very luminous.

Four CO lines, from $J_{\rm up} = 4$ to $J_{\rm up} = 7$, are detected in the FTS spectrum of this galaxy, all of them in the SLW range. In the modeling we included ground-based observations of the \CO{1}, $J = 2 - 1$, and $J = 3 - 2$ lines (see Section \ref{ss:groundCO}).
The ground-based data shows three kinematical components (see Figure \ref{fig:iram_fit_ngc3982} and Section \ref{ss:groundCO_fitting}). The width of the middle component is 100\,km\,s$^{-1}$. The other two components are probably produced by a rotating disk. The difference in the relative intensities of the redshifted and blueshifted disk components from the $J = 2 - 1$ to the $J = 3 - 2$ transition are presumably due to pointing uncertainty and the smaller beam size of the \CO2\ observation. Mapping observations of these transitions would be needed to confirm this interpretation.

The best-fit model for NGC~3982 has a higher density and lower temperature than the rest of the galaxies (see Table \ref{tab:radexfit} and Figure \ref{fig:radex_model}).
No CO lines with $J_{\rm up}>7$ are detected, so our fit could be biased toward low kinetic temperatures. 
However, the CO $J=4-3$\slash CO $J=7-6$ ratio in NGC~3982 is more than a factor of two higher than in the other galaxies in the sample. This implies a lower kinetic temperature for a fixed density and, therefore, the estimated lower temperature is likely real and not a consequence of the limited sensitivity of the \FTS\ spectrum.

\subsubsection{NGC~4388}

NGC~4388 is a nearby edge-on spiral galaxy located at 18.2\,Mpc. 
A weak broad H$\alpha$ component has been reported by \citet{Shields1996} and it is classified as Seyfert 1.9 by \citet{Ho1997}. However \citet{Ruiz1994} did not find evidence for broad components in the near-IR Pa$\beta$ line. A hidden broad line region (HBLR) is detected in optical polarized light \citep{Young1996}. Both the mid-IR [\ion{Ne}{5}] and [\ion{O}{4}] high-ionization emission lines are detected in the \Spitzer\slash IRS spectrum of this galaxy \citep{Tommasin2010}. The absorption corrected hard X-ray luminosity of the AGN is 3.0$\times$10$^{41}$\,erg\,s$^{-1}$ \citep{Cappi2006}. A radio jet is observed in this galaxy \citep{Falcke1998}.

\begin{figure*}
\centering
\includegraphics[width=.9\columnwidth]{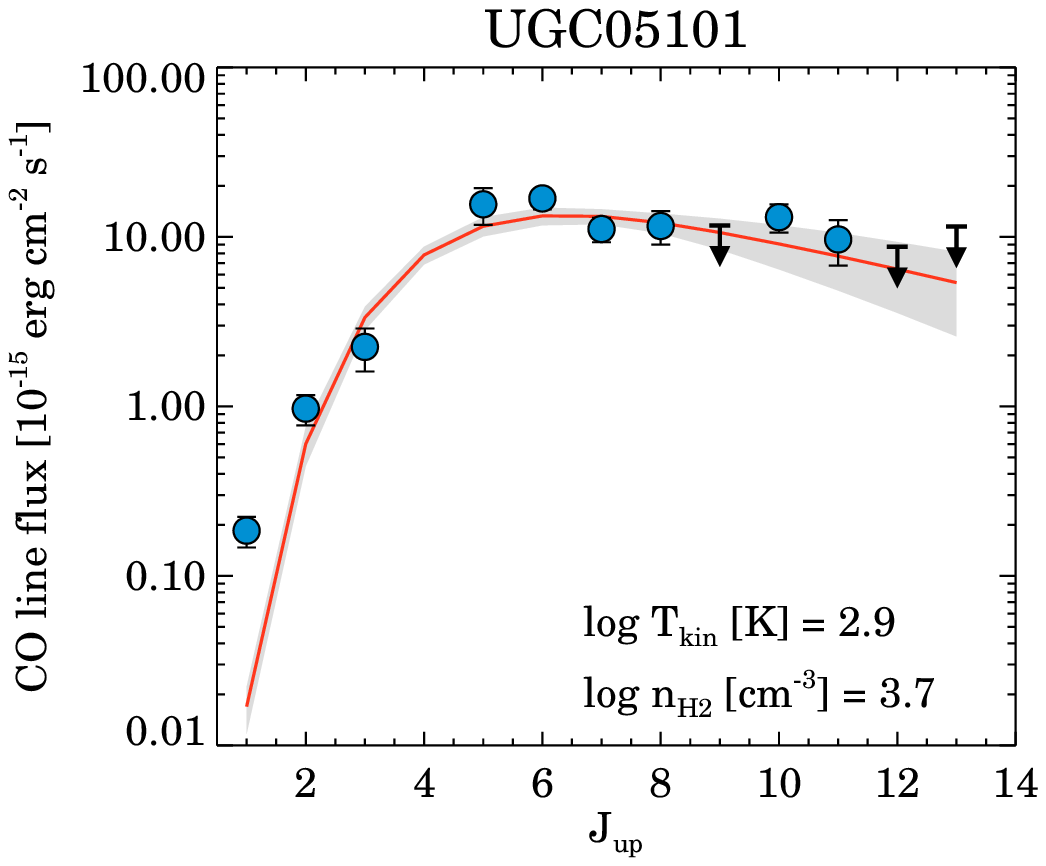}
\includegraphics[width=.9\columnwidth]{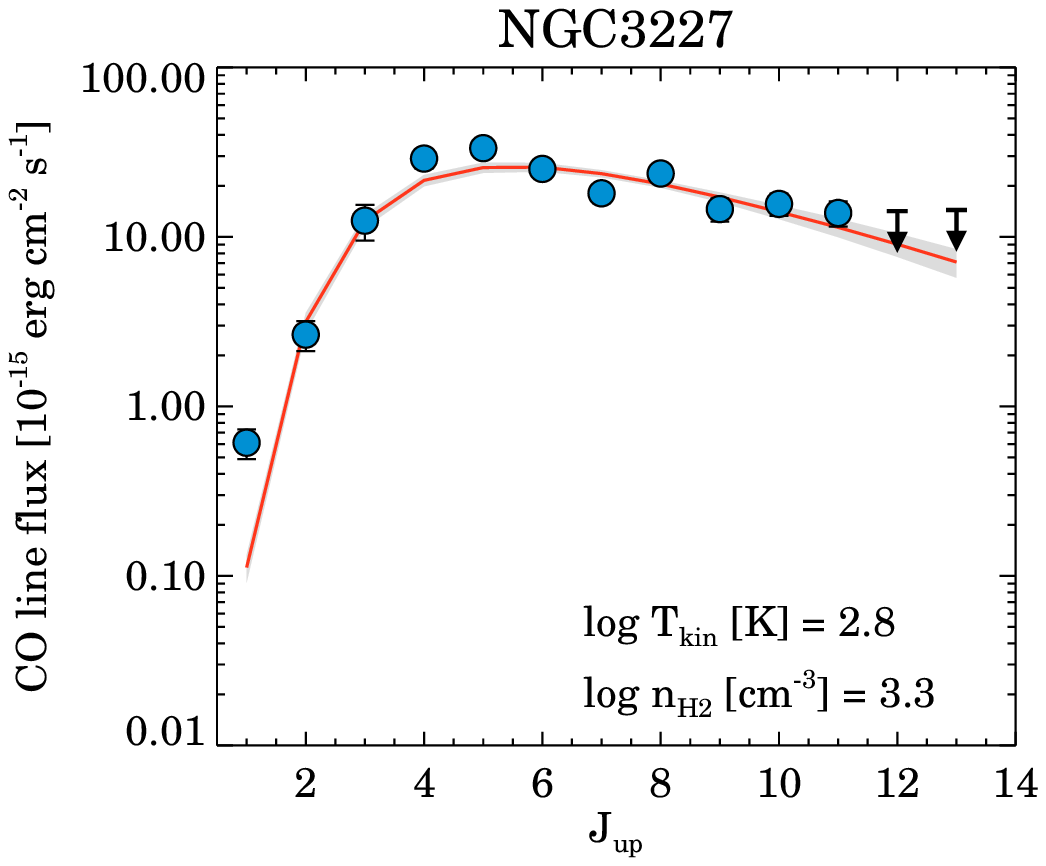}
\includegraphics[width=.9\columnwidth]{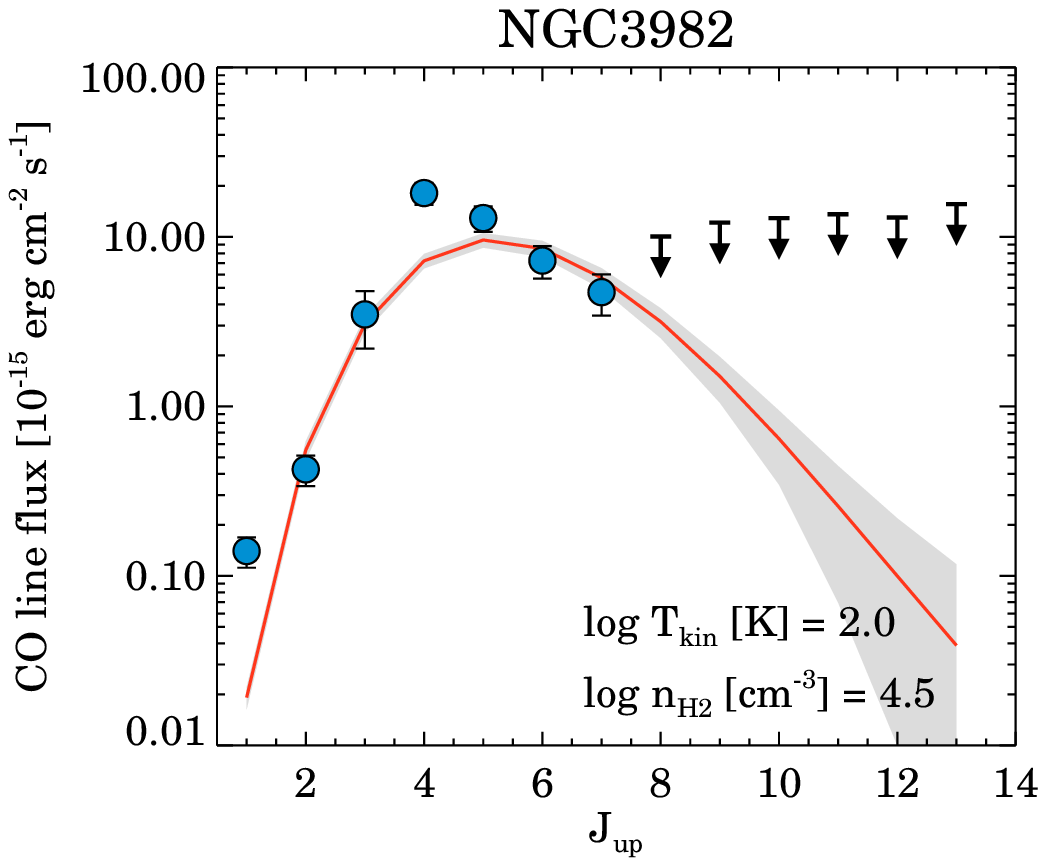}
\includegraphics[width=.9\columnwidth]{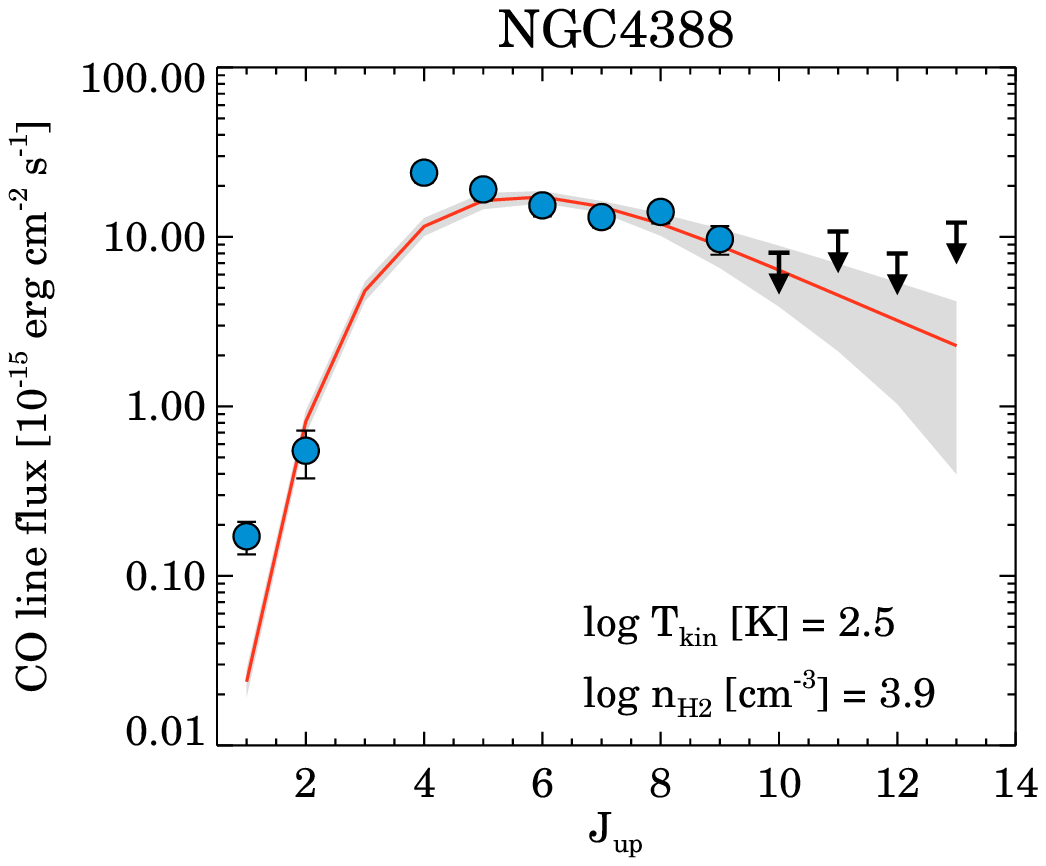}
\includegraphics[width=.9\columnwidth]{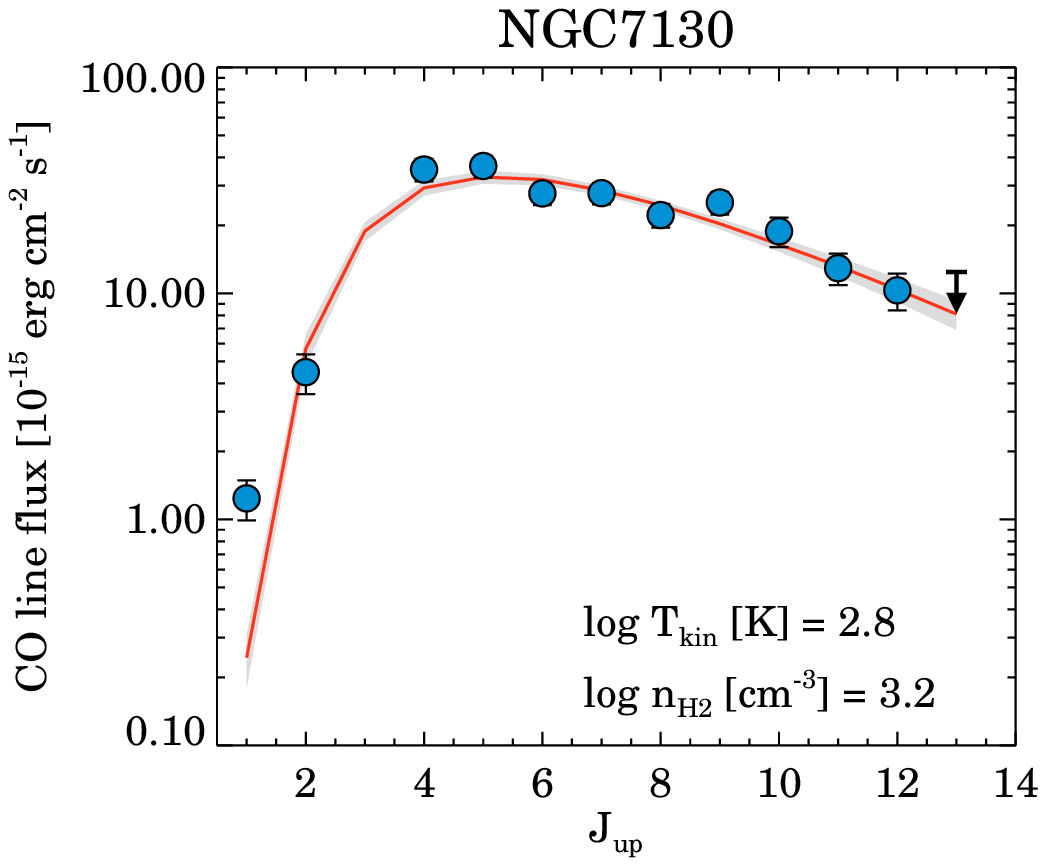}
\includegraphics[width=.9\columnwidth]{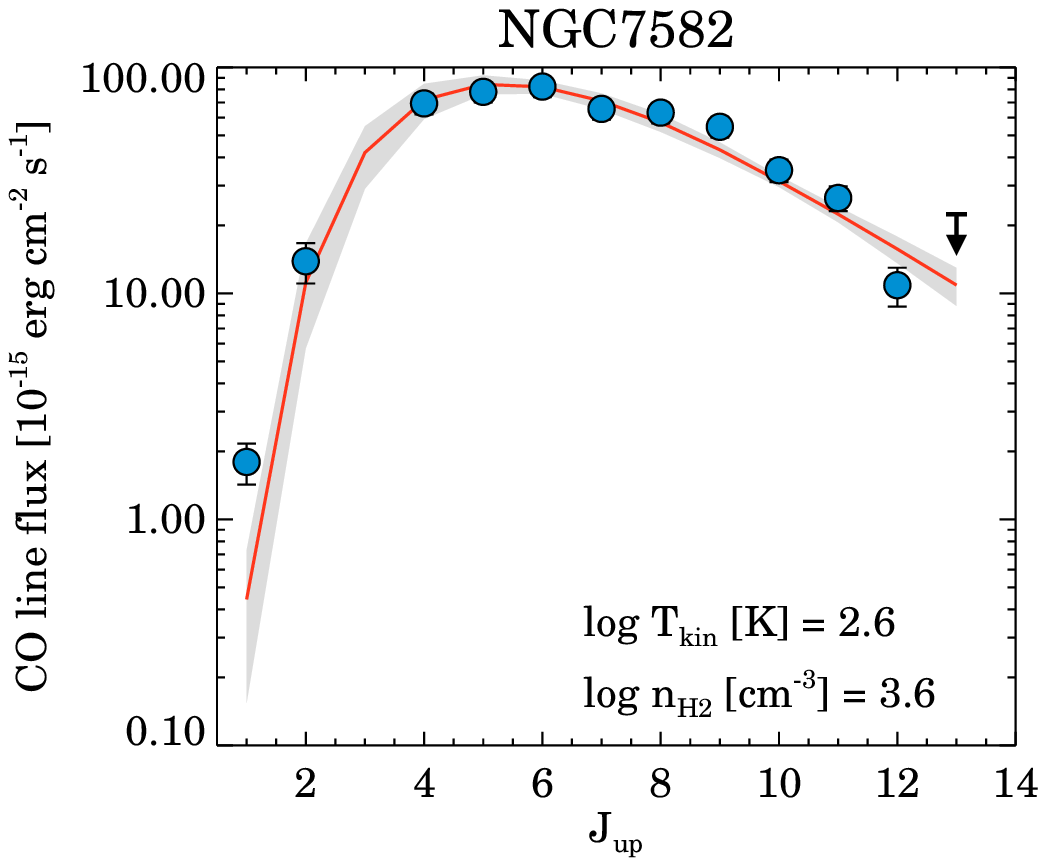}
\hspace{6cm}
\caption{CO SLED (blue circles) and best-fit model (red line) for our galaxies.\label{fig:radex_model} The shaded region is the 1$\sigma$ uncertainty in the best-fit model calculated from the Monte Carlo simulations (see Section \ref{s:rad_trans}).}
\end{figure*}

We detect six CO lines from $J_{\rm up}=4$ to 9 in the SLW \FTS\ spectrum. We obtained ground-based measurements of the \CO{1} and $J=2-1$ fluxes from the literature (\citealt{Pappalardo2012}, see Section \ref{ss:groundCO}). For the line width we assumed 125\,km\,s$^{-1}$ \citep{Vila-Vilaro1998}. The best-fit model is shown in Figure \ref{fig:radex_model}.

\subsubsection{NGC~7130}

NGC~7130 is a luminous IR galaxy ($\Lir = 10^{11.35}\,\Lsun$) located at 66\,Mpc \citep{SandersRBGS}. Its nuclear activity is classified as Seyfert 2 \citep{Yuan2010}. It is a face-on spiral galaxy with a disturbed morphology. Faint tidal tails are observed, although there are no nearby companions. It hosts a Compton-thick AGN with an estimated intrinsic hard X-ray luminosity of 8.3$\times$10$^{40}$\,erg\,s$^{-1}$ \citep{Levenson2005}. \Spitzer\slash IRS observations show the detection of the [\ion{Ne}{5}] and [\ion{O}{4}] high-ionization emission lines \citep{Tommasin2010}.
In addition to the powerful AGN, a strong burst of star-formation is taking place in the nuclear region of this galaxy \citep{Gonzalez-Delgado2001,Diaz-Santos2010}.

In the \FTS\ spectrum of NGC~7130 we measured nine CO lines from $J_{\rm up} =4$ to 12. We fitted the lines observed with \FTS\ together with the \CO{1} and \CO{2} fluxes obtained from the literature \citep{Albrecht2007}. The CO line width is $\sim$90\,km\,s$^{-1}$ \citep{Albrecht2007}. 

The \CO{10} line was excluded from the fit because it could be blended with the H$_2$O\,3$_{12}$--2$_{21}$ line. However the best-fit model explains the observed flux of the \CO{10} line, and therefore the contribution of the water line is likely small (see Figure \ref{fig:radex_model}).

\subsubsection{NGC~7582}

NGC~7582 is a Seyfert galaxy at a distance of 20.6\,Mpc. It is a member of the interacting Grus Quartet \citep{deVacouleurs1975}.
The nuclear activity is classified as Type 2 because no broad lines are detected in the optical \citep{Heisler1997}. However, \citet{Reunanen2003} reported a broad component in the near-IR Br$\gamma$ line. The hard X-ray luminosity of the AGN is 2.0$\times$10$^{41}$\,erg\,s$^{-1}$ \citep{Piconcelli2007}.
The nuclear region of NGC~7582 harbors young star-forming regions that dominate the ionization of the nuclear gas \citep{Riffel2009}.
In the mid-IR \Spitzer\slash IRS spectrum of NGC~7582 the [\ion{Ne}{5}] and [\ion{O}{4}] high-ionization emission lines are detected \citep{Tommasin2010}.

The CO lines from $J_{\rm up} =4$ to 12 are detected in the \FTS\ data. We added to the CO SLED ground-based observations of the \CO{1} and $J = 2-1$ lines \citep{Aalto1995}. 
For the line width we used $\sim$200\,km\,s$^{-1}$ \citep{Baan2008}.

The nuclear far-IR emission of this galaxy is only partially resolved at the resolution of \Herschel, so the effects of the different \FTS\ beam sizes are minimized. Therefore, we fitted all the CO lines simultaneously (Figure \ref{fig:radex_model}).

\subsection{Cold Molecular Gas}\label{s:co_cold}

From Figure \ref{fig:radex_model} we see that the best-fit models for the warm molecular gas components ($T_{\rm kin}\sim$500\,K; see Section \ref{s:co_warm}) underpredict the flux of the \CO1\ line. The predicted fluxes are 10--25\,\% of the observed values. This \CO1\ excess is explained if a cold gas component is also present in these galaxies.

The beam-averaged H$_2$ column density of the cold component can be estimated using the CO-to-H$_2$ conversion factor $N_{\rm H_2}$(cm$^{-2}$) = $0.5\times 10^{20} I_{\rm CO J=1-0}$\,(K\,km\,s$^{-1}$) revised for active galaxies \citep{Downes1998}. Substituting in Equation \ref{eq:masa_mol} we obtain: \\$M_{\rm mol}$(\Msun) $\sim7 \times 10^{20} F_{\rm CO J=1-0}$(erg\,cm$^{-2}$\,s$^{-1}$)$\times$ $(d$(Mpc)$)^2$.
We subtracted the contribution of the warm component to the \CO1\ line for the six galaxies modeled. For the rest of the galaxies this contribution to the \CO1\ emission introduces an extra $\sim$25\,\% uncertainty in the cold molecular mass.
The calculated masses are listed in Table \ref{tab:gasprop}; they range from 10$^{7.6}$ to 10$^{9.5}$\,\Msun.

\begin{deluxetable*}{lcccccccccc}
\tablewidth{0pt}
\tabletypesize{\small}
\tablecaption{ \label{tab:gasprop} Molecular Gas Properties} 
\tablehead{\colhead{Galaxy} & \colhead{${\log M_{\rm warm}}^a$} & \colhead{${\log L_{\rm CO}}^b$} & \colhead{${\log L_{\rm H_2 S(1)}}^c$} & \colhead{${L_{\rm CO} \slash M_{\rm warm}}^d$} & \colhead{${\log M_{\rm cold}}^e$} & \colhead{${\log M_{\rm C}}^f$} & \colhead{${x_{\rm C}}^g$} \\
& (\Msun) &  (\Lsun) & (\Lsun) & (\Lsun\slash\Msun) &  (\Msun) & (\Msun)}
\startdata
NGC~1056 & \nodata  & \nodata & 6.1 & \nodata & 8.3   & 4.6 & 5.6$\times$10$^{-5}$ \\
UGC~05101 & 7.7 & 7.9  & 8.0 & 1.4        & 9.5 & 6.4  & 1.5$\times$10$^{-4}$\\
NGC~3227  & 6.2 & 6.1  & 6.3 & 0.6        & 8.0  & 4.7  & 4.1$\times$10$^{-4}$\\
NGC~3982  & 5.9 & 5.7  & 5.6 & 0.6        & 7.6  & 4.4  & 1.3$\times$10$^{-4}$\\
NGC~4051 & \nodata  & \nodata & 5.8 & \nodata & 8.0  & 4.0 & 2.5$\times$10$^{-5}$\\
NGC~4151 & \nodata  & \nodata & 5.4 & \nodata & \nodata & 3.8 & \nodata \\
NGC~4388  & 6.0 & 6.0 & 6.3 & 1.0        & 7.6  & 4.7  & 8.4$\times$10$^{-4}$\\
IC~3639 & \nodata &  \nodata & 6.4 & \nodata & 8.5  & 4.8 & 4.8$\times$10$^{-5}$\\
NGC~7130  & 7.8 & 7.5  & 7.1 & 0.4        & 9.6 & 6.0 & 1.7$\times$10$^{-4}$\\
NGC~7172 & \nodata & \nodata & 6.8 & \nodata & 9.0  & 5.8 & 3.6$\times$10$^{-4}$\\
NGC~7582  & 7.0 & 6.9  & 6.7 & 0.6        & 8.7  & 5.1 & 1.8$\times$10$^{-4}$
\enddata
\tablecomments{For the warm and cold molecular masses ($M_{\rm warm}$ and $M_{\rm cold}$) listed in this table we assumed $M_{\rm mol}=1.4 M_{\rm H_2}$ to account for He mass. \\ 
$^{(a)}$ Mass of the warm molecular gas traced by the mid-$J$ CO lines (see Section \ref{s:co_warm}). 
$^{(b)}$ Integrated luminosity of the CO lines from $J_{\rm up}=4$ to $J_{\rm up}=12$ of the best-fit RADEX model (see also Table \ref{tab:radexfit}).
$^{(c)}$ Extinction corrected luminosity of the H$_2$ \hbox{0 -- 0} S(1) line at 17.03\,\micron.
$^{(d)}$ Warm molecular gas cooling rate per unit gas mass due to the CO transitions from $J_{\rm up}=4$ to $J_{\rm up}=12$.
$^{(e)}$ Mass of the cold molecular gas derived from the \CO1\ line flux (see Section \ref{s:co_cold}). $^{(g)}$ Neutral carbon mass in the cold molecular clouds (Section \ref{s:neutral_carbon}). 
$^{(f)}$ Neutral carbon abundance with respect to H$_2$.}
\end{deluxetable*}

\subsection{Warm Molecular Gas}\label{s:co_warm}

From the RADEX modeling of the low- and mid-$J$ CO lines, we find that the CO SLEDs of five out of the six galaxies analyzed (all except NGC~3982) can be fitted with a warm gas component with comparable physical properties, $n_{\rm H_2}\!\sim$10$^{3.2}$--10$^{3.9}$\,cm$^{-3}$ and $T_{\rm kin}\!\sim$300--800\,K (see Table \ref{tab:radexfit}).

Similar analysis of the mid-$J$ CO SLED have been presented for Arp~220, M~82, and NGC~1068 \citep{Rangwala2011, Kamenetzky2012, Spinoglio2012}. In these galaxies two or more CO components are fitted to the SLED; therefore we compare our results with their warm gas component. The gas density and temperature in our galaxies are close to those of the starburst M~82 ($n_{\rm H_2}=10^{3.4}$\,cm$^{-3}$ and $T_{\rm kin}=500$\,K). In the ultraluminous IR galaxy (ULIRG) Arp~220 the warm molecular gas has higher temperature, $T_{\rm kin}=1300$\,K, and slightly lower density, $n_{\rm H_2}=10^{3.2}$\,cm$^{-3}$, although within the uncertainties, both parameters are comparable to the values of our galaxies. In contrast, the warm molecular gas traced by the mid-$J$ CO lines in NGC~1068 has a lower temperature, 90\,K, and higher density, 10$^{4.6}$\,cm$^{-3}$.

We use Equation \ref{eq:masa_mol} to calculate the molecular mass. The resulting masses are listed in Table \ref{tab:gasprop}. They range from 10$^6$ to 10$^8$\,\Msun. These masses are lower than the warm molecular component observed in Arp~220, 10$^{8.7}$\,\Msun, but resemble those observed in M~82 and NGC~1068, 10$^{6.2}$\,\Msun\ and 10$^{7.4}$\,\Msun, respectively.
For these six galaxies the mass of the cold molecular gas (Section \ref{s:co_cold}) is between 40 and 120 times higher than that of the warm component, whereas in Arp~220 and M~82 the cold to warm molecular gas mass ratio is smaller, $\sim$10.

Warm molecular gas is also traced by the mid-IR rotational H$_2$ lines. The gas temperatures derived from the lowest H$_2$ rotational transitions are 100--1000\,K \citep{Rigopoulou1999,Roussel07} and are similar to the temperature range obtained from our radiative transfer analysis. Therefore, the mid-$J$ CO lines and the rotational H$_2$ lines might trace the same warm molecular clouds.

The rotational H$_2$~$0-0$~S(1) line at 17.03\,\micron\ is expected to account for 15--80\,\%\ of the total H$_2$ emission arising from the warm molecular gas (based on the \citealt{Shaw2005} models) depending on the gas temperature and density. Indeed, the H$_2$~S(1) is usually the brightest H$_2$ line in the mid-IR \Spitzer\slash IRS spectra of galaxies (e.g., \citealt{Roussel07}).
For most of the galaxies in our sample H$_2$ S(1) fluxes were presented by \citet{Tommasin08, Tommasin2010}\footnote{For UGC~05101 the H$_2$ S(1) flux was published by \citet{Farrah07}. For NGC~3227 and NGC~4151, S. Tommasin provided us with the H$_2$ S(1) fluxes measured from the high spectral resolution \Spitzer\slash IRS spectra following the method described in \citet{Tommasin2010}. The H$_2$ S(1) fluxes are (26.8 $\pm$ 0.2)$\times$10$^{-14}$ and (5.44 $\pm$ 0.82)$\times$10$^{-14}$\,erg\,cm$^{-2}$\,s$^{-1}$ for NGC~3227 and NGC~4151, respectively.}.
To correct for dust extinction first we estimated $A_{\rm K}$ using the relation $A_{\rm K}\slash \tau_{9.7}=1.48$ \citep{McClure2009}, where $\tau_{9.7}$ is the optical depth of the 9.7\,\micron\ silicate absorption feature. The $\tau_{9.7}$ values of our galaxies were published by \citet{Wu2009}. Then, assuming $A_{\rm 17\mu m} \sim 0.4\times A_{\rm K}$ \citep{McClure2009}, we corrected the H$_2$ S(1) fluxes.
In Table \ref{tab:gasprop} we compare the extinction corrected luminosity of this line with the integrated luminosity of the mid-$J$ CO lines ($J_{\rm up}=4$ to $J_{\rm up}=12$, those in the spectral range observed by \FTS). The total H$_2$ luminosity is 0.1--0.8\,dex higher than the H$_2$ S(1) luminosity depending on the exact gas conditions. So, in general, the H$_2$ cooling is higher but comparable to the mid-$J$ CO cooling in these galaxies.

Table \ref{tab:gasprop} also shows the ratio between the integrated luminosity of the mid-$J$ CO lines and the molecular gas mass ($L_{\rm CO}$\slash $M_{\rm mol}$). This ratio is between 0.4 and 1.3\,\Lsun\slash\Msun\ in our galaxies, similar to the CO cooling ratio measured in the warm CO component of Arp~220 (0.4\,\Lsun\slash\Msun\ \citealt{Rangwala2011}). For the gas conditions derived from the modeling, the expected H$_2$ cooling ratio is $\sim$0.5--10\,\Lsun\slash\Msun\ \citep{LeBourlot1999}. These values agree with the observed rotational H$_2$ cooling; thus this supports the idea that the mid-IR rotational H$_2$ lines and the mid-$J$ CO lines originate in the same warm molecular gas. This agreement also indicates that the CO abundance ($x_{\rm CO}=3\times10^{-4}$) used to calculate the $M_{\rm mol}$ is reasonable for the warm molecular gas in these active galaxies.

\subsection{Warm CO Heating Source}\label{ss:warm_co_heating}

\begin{figure}[ht]
\centering
\includegraphics[width=0.95\columnwidth]{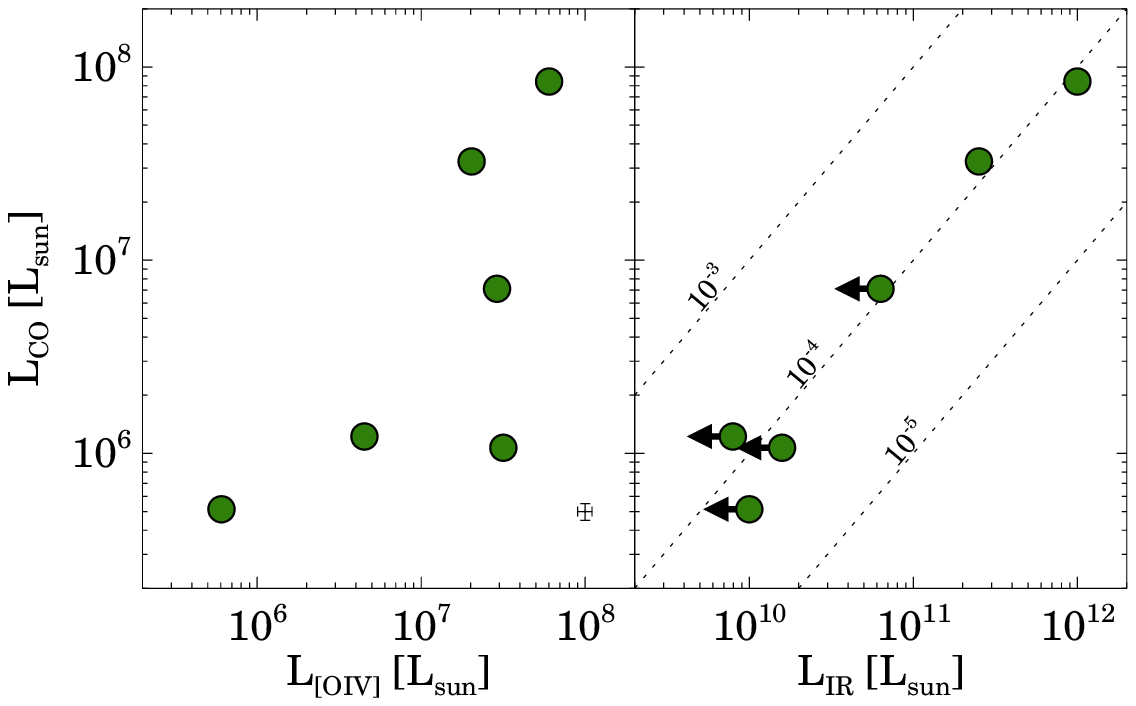}
\caption{Mid-$J$ CO luminosity vs. [\ion{O}{4}]25.9\,\micron\ luminosity (left panel) and vs. IR luminosity (right panel). The [\ion{O}{4}]25.9\,\micron\ luminosity is a proxy of the AGN luminosity in Seyfert galaxies whereas the IR luminosity traces the star-formation rate. The dashed lines in the right panel show the $L_{\rm CO}$\slash $L_{\rm IR}$ ratio. The [\ion{O}{4}]25.9\,\micron\ fluxes are from the \citet{Pereira2010c} compilation. \label{fig:co_lum}}
\end{figure}

Our radiative transfer modeling shows that warm molecular gas contributes significantly to the CO luminosity in these Seyfert galaxies. To determine the heating source of the warm gas in Figure \ref{fig:co_lum} we compare the mid-$J$ CO luminosity with the [\ion{O}{4}]25.9\,\micron\ luminosity, a proxy of the AGN luminosity \citep{Melendez2008}, and the IR luminosity\footnote{The $L_{\rm IR}$ values in Table \ref{tab:sample} are calculated from \textit{IRAS} fluxes. \textit{IRAS} beams are much larger than the \FTS\ beam, thus for the galaxies with the larger apparent sizes (see Figure \ref{fig:ftsfootprint}) we consider this $L_{\rm IR}$ as an upper limit.}, which is correlated with the star-formation rate \citep{Kennicutt1998}. There is no clear correlation between the AGN luminosity and the CO luminosity (left panel). For $L_{\rm [O\,IV]}\sim 5\times10^{7}$\,\Lsun, the CO luminosity differs two orders of magnitude from galaxy to galaxy. On the other hand, the $L_{\rm CO}$ seems consistent with a $L_{\rm CO}$\slash $L_{\rm IR}$ ratio about 10$^{-4}$. That is, the mid-$J$ CO emission is more likely related to the star-formation activity than to the AGN luminosity in our sample of Seyfert galaxies.

PDRs are expected in star-forming regions where UV photons from young stars heat the ISM; therefore we compared the observed CO SLEDs with the PDR models of \citet{Wolfire2010}. However, we find that no single PDR model reproduces the observed CO SLED. In general the best-fit PDR models underpredict the fluxes of the CO lines with $J_{\rm up}>8$ lines. Conversely, XDR models \citep{Meijerink2006} reproduce better the higher $J$ CO lines and slightly underpredict those with $J_{\rm up}<6$. Consequently we cannot rule out an XDR contribution to the higher $J$ CO emission in our galaxies.

\begin{figure}[t]
\centering
\includegraphics[width=0.95\columnwidth]{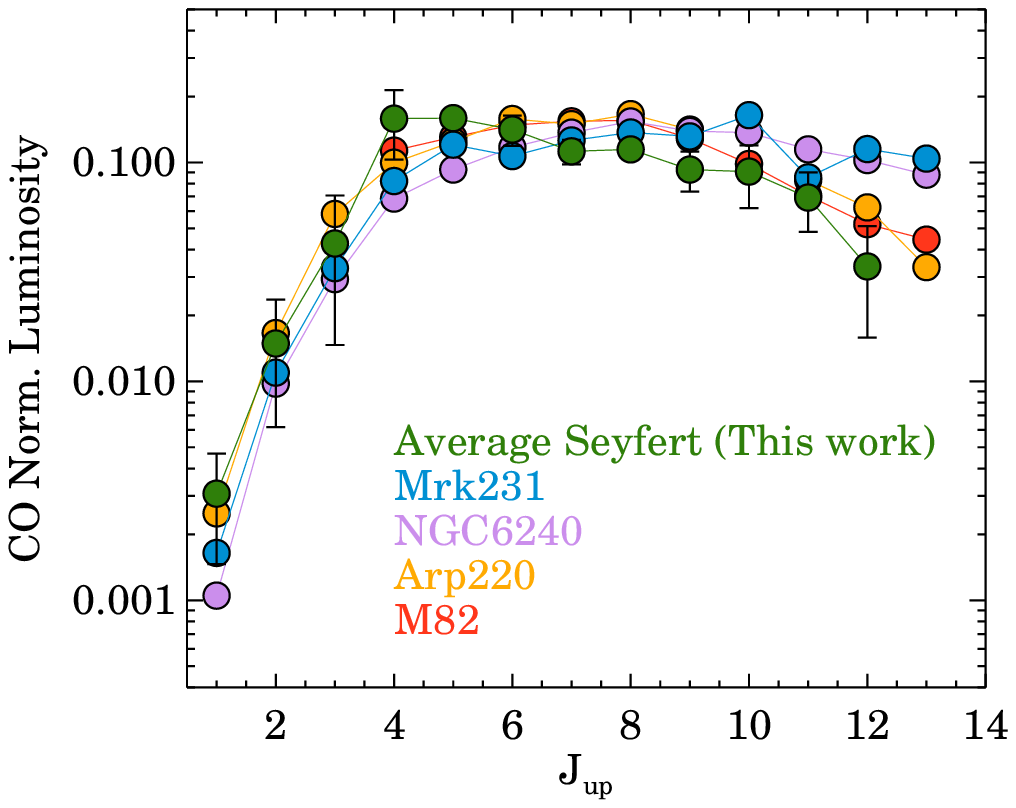}
\caption{CO SLEDs normalized to the integrated CO luminosity (from $J_{\rm up}=1$ to $J_{\rm up}=12$). We compare the average CO SLED of our Seyfert galaxies with those of Mrk~231 \citep{vanderWerf2010}, NGC~6240 \citep{Meijerink2013}, Arp~220 \citep{Rangwala2011}, and M~82 \citep{Kamenetzky2012}.
\label{fig:co_sled_compara}}
\end{figure}

In Figure \ref{fig:co_sled_compara} we compare the average CO SLED of our Seyfert galaxies with those published for other galaxies.
The luminosity normalized CO SLEDs are visibly very similar up to $J_{\rm up}=11$. Only for $J_{\rm up}>11$, the CO emission in Mrk~231 and NGC~6240 is approximately a factor of 3 brighter than in the rest of galaxies. For the Seyfert 1 Mrk~231 an XDR is proposed to explain the CO $J_{\rm up}>10$ emission and two PDR to model the lower-$J$ CO emission \citep{vanderWerf2010}, whereas for the merger galaxy NGC~6240 shock excitation can explain its complete CO SLED \citep{Meijerink2013}. For Arp~220 PDR, XDR, and cosmic rays heating mechanisms are ruled out and mechanical energy from supernovae is suggested as the heating source \citep{Rangwala2011}. For M~82 a combination of PDR and shocks is required to explain the observed CO emission \citep{Kamenetzky2012}. That is, the mid-$J$ CO emission of galaxies, at least up to $J_{\rm up}\sim11$, seems to be produced by warm molecular gas with similar average physical conditions, although it can be heated by different combinations of mechanisms (PDR, XDR, cosmic rays, and shocks).

\section{[\ion{C}{1}] 492 and 809\,GHz Emission}\label{s:neutral_carbon}

PDR models predict that the two [\ion{C}{1}] lines at 492 and 809\,GHz are produced in the transition region from C$^+$ to CO that occurs in a relatively thin layer within the molecular clouds \citep{Kaufman1999, Bolatto1999}.
However some observational evidence suggests that neutral carbon and CO can coexist in the same volume (see \citealt{Papadopoulos2004} and references therein). The latter can be achieved due to non chemical equilibrium between C$^0$ and CO \citep{Papadopoulos2004,Glover2010} or turbulent diffusion of the neutral carbon layer into the interiors of the molecular clouds \citep{Xie1995}. Moreover, cosmic rays and X-ray radiation also increase the C$^0$\slash CO abundance ratio throughout the clouds \citep{Papadopoulos2004,Meijerink2006}.

For five of the galaxies in our sample we detect both [\ion{C}{1}] fine structure emission lines at 492 and 809\,GHz. For the rest we detect only the [\ion{C}{1}] line at 809\,GHz (see Table \ref{tab:lines}). This is in part because the 492\,GHz emission line lies in a noisy part of the SLW spectra, but also because it seems to be weaker than the 809\,GHz line (at least in the galaxies with both [\ion{C}{1}] lines detected).

To investigate the origin of the [\ion{C}{1}] emission in these galaxies we used RADEX to calculate the fluxes of the \Cia\ and 809\,GHz lines for wide range of physical conditions ($n_{\rm H_2}$=10--10$^8$\,cm$^{-3}$, $T$=10--1000\,K, and $N_{\rm C}$\slash$\Delta v$=10$^{12}$--10$^{18}$ cm$^{-2}$\slash(km\,s$^{-1}$)). We considered C$^0$ collisions with H$_2$ using the collisional rate coefficients of \citet{Schroder1991}.

\begin{figure}
\centering
\includegraphics[width=0.95\columnwidth]{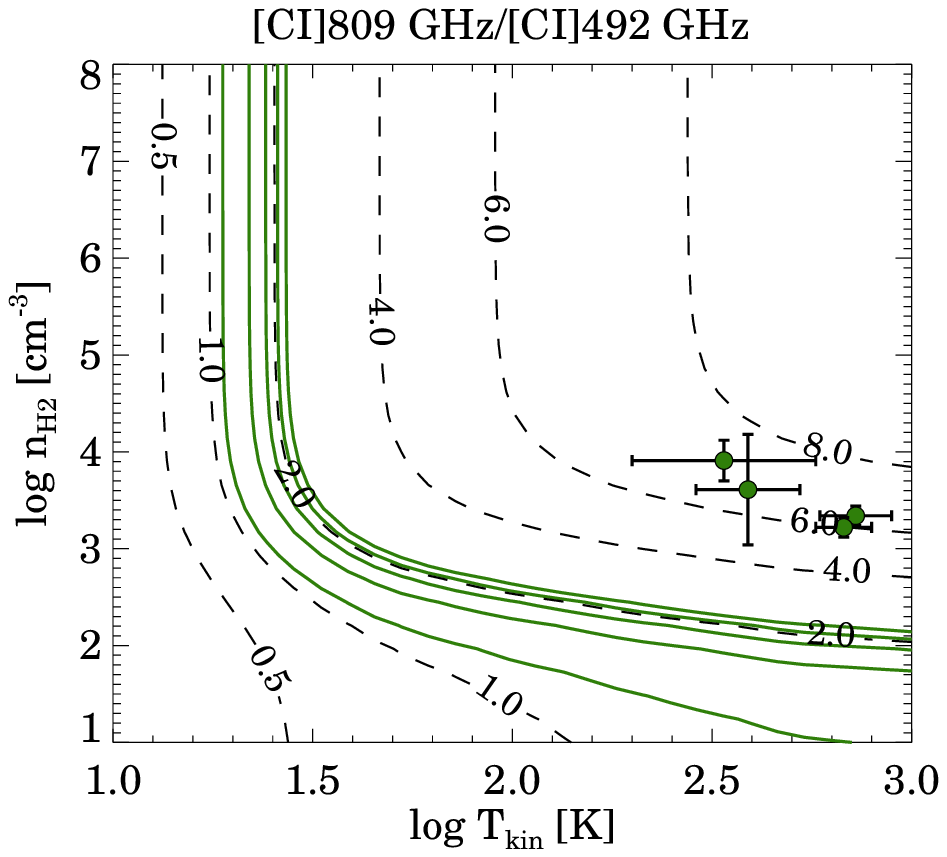}
\caption{The dashed black lines are the \Cib\slash\Cia\ ratio (fluxes in erg\,cm$^{-2}$\,s$^{-1}$\,sr$^{-1}$) as a function of the H$_2$ density ($n_{\rm H_2}$) and kinetic temperature ($T_{\rm kin}$) in the optically thin limit. The solid green lines are the ratios measured in our sample of galaxies. The green circles represent the molecular gas conditions derived from the mid-$J$ CO lines (see Section \ref{s:rad_transfer}). UGC~05101 is not included in this figure because only the \Cib\ line is detected in this galaxy.\label{fig:c1ratio}}
\end{figure}

In Figure \ref{fig:c1ratio} we plot the expected \Cib\slash\Cia\ ratio as a function of the molecular hydrogen density and kinetic temperature for optically thin emission. The solid lines represent the observed ratios for the five galaxies with both lines detected. They span a narrow range in our sample, from 1.2 to 2.2.
These \Cib\slash\Cia\ ratios indicate a kinetic temperature about 19--27\,K 
if the gas density is higher than the critical densities of the neutral carbon lines ($n_{\rm H_2}> 10^4$\,cm$^{-3}$). For lower gas densities, the temperature is not well constrained since this ratio would be compatible with that of higher temperature ($T_{\rm kin}\!>$50\,K) diffuse clouds (lower right corner in Figure \ref{fig:c1ratio}).

To test if the [\ion{C}{1}] emission arises from the warm molecular gas component traced by the mid-$J$ CO lines, we plotted the derived physical conditions of the warm gas (see Section \ref{s:rad_transfer}) in Figure \ref{fig:c1ratio}. Under these conditions the expected \Cib\slash\Cia\ ratio is 6--8, more than three times the observed ratio, so the bulk of the neutral carbon emission is not likely to be produced in the warm molecular gas component.

Instead, the [\ion{C}{1}] emission can originate in the cold gas component traced by the CO $J=1-0$ line (see Section \ref{s:co_cold}). Actually there is a good correlation between the C$^0$ and \CO1\ intensity in our Galaxy \citep{Ikeda2002,Ojha2001} and nearby galaxies \citep{Wilson1997}. However, the ratio between the CO and [\ion{C}{1}] emissions depends on the specific physical conditions (UV flux, cosmic ray rate, turbulence, etc.) in the molecular clouds \citep{Israel2002,Papadopoulos2004}. 

We used the RADEX grid of models to calculate the C$^0$ mass and column density in these cold molecular clouds. We assumed that the neutral carbon emission is optically thin and thermalized ($n_{\rm H_2}>10^4$\,cm$^{-2}$). From the \Cib\slash\Cia\ ratio we estimated the gas temperature (see Figure \ref{fig:c1ratio}). When the \Cia\ line is not detected we assumed $T_{\rm kin}=23$\,K (the median $T_{\rm kin}$ for the galaxies with both [\ion{C}{1}] lines detected). Then we compared the observed \Cib\ luminosity with the predicted value to estimate the C$^0$ column densities and masses. The derived C$^0$ masses range from 10$^{3.8}$--10$^{6.3}$\,\Msun\ (Table \ref{tab:gasprop}). In Section \ref{s:co_cold} we calculated the mass of the cold molecular component; therefore using that the relative weight of C$^0$ and H$_2$ is 6, we obtain that the C$^0$ abundance relative to H$_2$ in these galaxies is $\sim$3$\times 10^{-5}$--8$\times 10^{-4}$. Consequently, the C$^0$\slash CO abundance ratio ranges from 0.1 to 2 (assuming $x_{\rm CO}=3\times10^{-4}$). These C$^0$\slash CO abundance ratios are comparable to those measured in quiescent nearby galaxies \citep{Israel2002} and starbursts \citep{Rangwala2011,Kamenetzky2012}. However, it should be noted that these C$^0$ abundances rely on the CO-to-H$_2$ conversion factor that can vary up to a factor of $\sim$4 from galaxy to galaxy (e.g., \citealt{Downes1998}).

The C$^0$ abundance is expected to be enhanced in XDRs with respect to that in PDRs \citep{Meijerink2006}. For a constant gas density \citet{Meijerink2006} models predict higher \Cia\slash\CO1\ intensity ratios in XDRs than in PDRs. In these PDR models the \Cia\slash\CO1\ ratio ranges between 20 and 170, decreasing with increasing gas densities, whereas in XDR models this ratio is between 50 and 10$^5$. In most of our galaxies the \Cia\slash\CO1\ ratios are between 15 and 40, so they are compatible with PDR excitation.
Only in NGC~4388 the \Cia\slash\CO1\ ratio is higher, 85$\pm$20. This ratio can be explained by PDR models, but also by a high-density ($n=10^5$\,cm$^{-3}$) XDR model.
Therefore, in most of our sample the C$^0$ emission is consistent with that expected from PDRs, although for one galaxy XDR excitation is possible.

\section{Hydrogen Fluoride}\label{s:HF}

The HF $J = 1-0$ transition at 1232\,GHz has been recently detected in the \FTS\ spectra of several local galaxies: the ULIRGs Mrk~231 \citep{vanderWerf2010} and Arp~220 \citep{Rangwala2011}; the starburst M~82 \citep{Kamenetzky2012}; and the Seyfert 2 NGC~1068 \citep{Spinoglio2012}. This transition has also been observed in the Cloverleaf quasar at $z=2.46$ \citep{Monje2011}.
In our sample, we detect this line in UGC~05101 in absorption, and in NGC\,7130 in emission (see Table \ref{tab:lines_uncommon}).

Hydrogen fluoride molecules are rapidly formed by the reaction of F with molecular hydrogen when the latter becomes abundant, so it is a good proxy for molecular gas.

Most of the fluorine in molecular clouds is present as hydrogen fluoride, and in diffuse clouds HF can be several times more abundant than CO \citep{Neufeld2005}.

Owing to the high critical density of the HF $J = 1-0$ transition ($n_{\rm H_2}\sim$10$^{10}$\,cm$^{-3}$ at 50\,K; \citealt{Neufeld2005}), it is expected to be observed in absorption.
However, IR radiative pumping\footnote{Ground-state HF can be excited by 2.5\,\micron\ near-IR photons to its first vibrational level, that then can decay to excited rotational levels of the ground vibrational state.} can populate levels with $J \geq 1$, and thus it would be possible to observe the $J = 1-0$ transition in emission \citep{Neufeld2005}. Alternatively, HF can be formed in excited states that lead to the emission of the $J = 1-0$ line. This mechanism, chemical pumping, is possible in strong UV radiation fields where HF photodissociation and formation rates are high \citep{vanderTak2012}. In addition, since HF has a large dipole moment, collisions with electrons can excite its rotational levels \citep{vanderTak2012b}.

\subsection{UGC~05101}\label{ss:hf_ugc05101}

We detect the HF $J = 1-0$ transition in absorption in UGC~05101.
Assuming that this transition is optically thin, we calculated the HF column density using the following equation (see \citealt{Draine2011}):

\begin{equation}\label{eqn:coldens}
\small
N_l {\rm(cm^{-2})}= 9.33 \times 10^{5}\,\frac{W_\lambda {\rm(cm)}\,(\nu_{ul}\,{\rm (GHz)})^4}{A_{ul} {\rm (s^{-1})}}\,\frac{g_l}{g_u},
\end{equation}
where $N_l$ is the column density of the lower energy level, $A_{ul}$ the Einstein coefficient, $W_\lambda$ the equivalent width of the line, and $g_u$ and $g_l$ the degeneracy of the upper and lower levels, respectively. For the HF $J=1-0$ transition, $A_{10}$ = $2.42\times 10^{-2}$\,s$^{-1}$, $g_1$ = 3, and $g_0$ = 1, and $\nu_{10}$ = 1232.476\,GHz (from the Leiden Atomic and Molecular Database [LAMDA]; \citealt{Schoier2005}). In UGC~05101, the continuum at the frequency of the HF $J=1-0$ line is 5.4 $\pm$ 0.2\,Jy, so $W_\lambda = (3.8 \pm 0.9)\times 10^{-6}$\,cm. Substituting in Equation \ref{eqn:coldens} we obtain $N_0 = (1.1 \pm 0.3)\times 10^{14}$\,cm$^{-2}$.
This column density should be considered a lower limit because: (1) the measured equivalent width is a lower limit, since we assumed that all the observed continuum at 1232\,GHz illuminates the molecular clouds that produce the HF absorption and this might not be true; (2) we can only observe the HF in molecular clouds in front of the far-IR continuum source in our line of sight; and (3) we do not consider HF molecules in excited energy levels, although most of the HF molecules are expected to be in the ground energy level (see \citealt{Neufeld2005}).
\vspace{0.3cm}
\subsection{NGC~7130}

In NGC~7130 the HF $J=1-0$ transition is detected in emission. In this section we discuss three possible excitation mechanisms to explain the HF $J=1-0$  emission line in this galaxy.

First we calculated the column density of the upper level assuming that the HF emission is optically thin. This value is independent of the excitation mechanism.
Using RADEX, we verified that the optically thin approximation is valid for a wide range of physical conditions ($T_{\rm kin} = 5-250$\,K, $n_{\rm H_2} = 10-10^{10}$\,cm$^{-3}$, and $N_{\rm HF}=10^{8}-10^{13}$\,cm$^{-2}$ for $\Delta v$ = 90\,km\,s$^{-1}$). Therefore we can use following relation to calculate the column density of the $J=1$ level:
\begin{equation}\label{eqn:column_up}
<\!N_u\!> = \frac{4\pi}{\Omega}\frac{F_{ul}}{h \nu_{ul} A_{ul}},
\end{equation}
where $<\!N_u\!>$ is the beam averaged column density of the upper transition level, $\Omega$ the beam solid angle, $F_{ul}$ the line flux in erg\,cm$^{-2}$\,s$^{-1}$, and $h$ the Planck constant.
For the HF $J=1-0$ line the beam FWHM is 18\arcsec, thus $<\!N_1\!>=$(9.1 $\pm$ 1.2)$\times$10$^{10}$\,cm$^{-2}$. This column density is a lower limit to the total HF column density since we have only considered the molecules in the $J=1$ rotational level.

\subsubsection{IR Pumping}

\citet{Carroll1981} show that IR pumping can efficiently increase the population of ground state rotational levels when
\begin{equation}
\frac{f}{e^{{h\nu}\slash {kT}} - 1} > \frac{A_{J+1, J}}{A_{\rm \nu,\nu -1}},
\end{equation}
where $f$ is a factor to account for the geometric dilution and dust emissivity, $\nu$ the frequency of the vibrational transition, $T$ is the radiation field temperature, and $A_{J+1, J}$ and $A_{\rm \nu,\nu -1}$ are the Einstein coefficients for the rotational and vibrational transitions, respectively.
For the lowest vibrational transition $A=63.2$\,s$^{-1}$ and $h\nu\slash k=5756$\,K (LAMDA). Therefore assuming that the near-IR radiation seen by the HF molecules completely fills the sky ($f=1$), the required minimum radiation field temperature for efficient IR pumping is $T>730$\,K. This corresponds to an energy density $\nu u_{\rm \nu}\sim$5$\times 10^{-4}$\,erg\,cm$^{-3}$ at 2.5\,\micron, that is, a radiation field $>$10$^6$ times that in a PDR near to ($d=0.2$\,pc) an O star \citep{Draine2011}. If $f<1$ the radiation temperature, and energy density, would be higher. Consequently, if the HF molecules are excited by IR pumping in NGC~7130, they should be illuminated by a very intense IR radiation field that can be produced only by the AGN.

\subsubsection{Chemical Pumping}

To be efficient, chemical pumping requires that the number of HF molecules formed exceeds the number of spontaneous decays from the $J=1$ to the $J=0$ levels (see \citealt{vanderTak2012}). This condition can be expressed as
\begin{equation}\label{eqn:chemical}
R \frac{N_{\rm HF}}{n_{\rm HF}} \ge A_{10} N_{1},
\end{equation}
where $R$ is the formation rate, $N_{\rm HF}$ the total HF column density, $n_{\rm HF}$ the HF density, and $N_1$ the column density of the HF molecules in the $J$=1 level.
Assuming that HF formation and destruction are balanced:
\begin{equation}
R = k n_{\rm H_2} n_{\rm F} = n_{\rm HF}\zeta_d\chi,
\end{equation}
where $k = 2.86\times 10^{-12}$\,cm$^{3}$\,s$^{-1}$ is the HF formation rate at 50\,K, $\zeta_d = 1.17\times 10^{-10}$\,s$^{-1}$ the HF photodissociation rate for $\chi_{\rm UV}$, the standard interstellar UV radiation field of \citet{Draine1978} \citep{Neufeld2005}, $\chi$ the UV radiation field in units of $\chi_{\rm UV}$, and $n_{\rm HF}$ and $n_{\rm F}$ are the densities of hydrogen fluoride and atomic fluorine, respectively. Since $n_{\rm HF} + n_{\rm F} = 2{\cal A_{\rm F}} n_{\rm H_2}$, where ${\cal A}_{\rm F}=2.9\times 10^{-8}$ is the solar F abundance with respect to H \citep{Lodders2003}, Equation \ref{eqn:chemical} can be rewritten as
\begin{equation}
\frac{\zeta_d\chi}{A_{10}}\frac{2{\cal A}_{\rm F}kn_{\rm H_2}}{\zeta_d\chi + kn_{\rm H_2}}N_{\rm H_2} \ge N_1 .
\end{equation}

For the conditions in a PDR ($\chi=10^3$, $n_{\rm H_2}=10^4$\,cm$^{-3}$), the $N_1$ calculated before implies a $N_{\rm H_2} > 10^{24}$\,cm$^{-2}$. That is several orders of magnitude higher than the H$_2$ column density derived from the \CO1 transition (see Section \ref{s:co_cold}). Both higher gas densities and intenser radiation fields decrease the minimum H$_2$ column density. 
Only a combination of dense gas ($n_{\rm H_2}>10^7$\,cm$^{-3}$) and intense UV radiation fields ($\chi>10^{7}$) is compatible with the derived H$_2$ column density in NGC~7130. This intense UV radiation field suggests that the HF molecules should be close to the AGN.

\subsubsection{Collisional Excitation by Electrons}

The rotational levels of molecules are usually excited by collisions with H$_2$. However, for molecules with a large dipole moment, like HF, the rotational levels can be excited by collisions with electrons as well. In general, in the interior of molecular clouds, where most of the molecules are found, the electron abundance is low ($n_{\rm e}\slash n_{\rm H_2} < 10^{-6}$; see Figure 1 of \citealt{Meijerink2006}) and the collisions with electrons can be neglected. However, in some situations, like diffuse clouds \citep{Black1991}, and molecular clouds surfaces \citep{Wolfire2010}, or XDRs \citep{Meijerink2006}, the electron density is enhanced.
The majority of the free electrons in molecular clouds come from ionized carbon, with a small contribution from hydrogen ionized by cosmic rays \citep{Liszt2012}. Therefore, the electron abundance can be as high as $n_{\rm e}\slash n_{\rm H} \sim$ 1.6$\times$10$^{-4}$, that is, the gas-phase carbon abundance \citep{Sofia2004}.

The HF $J=1-0$ critical density for collisions with electrons at 50\,K is $n_{\rm e}$=1.7$\times$10$^{4}$\,cm$^{-3}$ (calculated using the LAMDA electron-HF collision strengths). Taking the highest $n_{\rm e}$ abundance calculated above, this critical density is reached when $n_{\rm H_2}$=10$^8$\,cm$^{-3}$. This is two orders of magnitude lower than the critical density for collisions with H$_2$, but it is still high. Hence, the HF $J=1-0$ emission line is thermalized only in high-density molecular clouds.

Similar to \citet{vanderTak2012b}, we used RADEX to model the HF emission taking into account collisions with both molecular hydrogen and electrons. We computed a grid of models for $n_{\rm H_2}=10^2$--10$^{11}$\,cm$^{-3}$, $T_{\rm kin}=10-250$\,K, and $n_{\rm e}\slash n_{\rm H_2}=10^{-4}$. We used the electron-HF collisional rate coefficients from LAMDA calculated by \citet{vanderTak2012b} using the Coulomb-Born approximation.
The HF $J=1-0$ line is optically thin for a wide range of physical conditions, so the line flux will be proportional to $N_{\rm HF}$ in the escape probability approximation. That is,
\begin{equation}\label{eq:HF_flux}
F {\rm (erg\,cm^{-2}\,s^{-1}\,sr^{-1})}={\cal K} N_{\rm HF}{\rm (cm^{-2})},
\end{equation}

where ${\cal K}$ depends on the density of the collisional partners, electrons and molecular hydrogen, and on the kinetic temperature.

For gas densities between $n_{\rm H_2}$=10$^{3}$--10$^{4}$\,cm$^{-3}$ and $T_{\rm kin}>50$\,K, $\cal K$ ranges between 10$^{-22}$ and 10$^{-21}$.
The HF $J=1-0$ flux in NGC~7130 is 1.4$\times$10$^{-6}$\,erg\,cm$^{-2}$\,s$^{-1}$\,sr$^{-1}$ (assuming the SLW beam size, 18\arcsec, as the source size). Accordingly, this corresponds to a beam averaged HF column density $>$10$^{15}$\,cm$^{-2}$. Given the F abundance, ${\cal A}_{\rm F}=2.9\times 10^{-8}$, the resulting $<\!N_{\rm H_2}\!>$ would be unrealistically high.
For lower $T_{\rm kin}$, ${\cal K}$ values are lower, $\sim$10$^{-24}$--10$^{-23}$, and the column density would be even higher.

In the high-density limit ($n_{\rm H_2}>$10$^{8}$\,cm$^{-3}$), where HF would be thermally excited, $\cal K$ is $\sim$10$^{-18}$. Therefore, for the observed HF flux in NGC~7130 we obtain  $<\!N_{\rm HF}\!>\sim$10$^{12}$\,cm$^{-2}$ and $<\!N_{\rm H_2}\!>\sim 10^{20}$\,cm$^{-2}$. It would imply $\sim 4\times 10^{7}$\,\Msun\ of very high density molecular gas where C is single ionized. This region would correspond to the dark molecular gas (that does not emit in CO), which represents about 30\% of the total gas in PDRs \citep{Wolfire2010}. However typical PDR densities are $n_{\rm H_2}\sim10^4$\,cm$^{-2}$, much lower than the high density gas where HF would be excited by electron collisions. These densities are reached in dense molecular cores, but there the UV field is shielded and ionized carbon is not expected. Therefore, the gas where the HF $J=1-0$ line originates does not seem to be related to the star-formation activity in this galaxy. On the contrary, X-rays from the AGN can keep a high ionization fraction over large column densities \citep{Gonzalez-Alfonso2013}. The importance of X-ray ionization in NGC~7130 is also supported by the detection of several OH$^+$ transitions in its FTS spectrum (see Section \ref{ss:OH}).

\subsubsection{HF Excitation in NGC~7130}

We discussed three possible excitation mechanisms to explain the HF $J=1-0$ emission in NGC~7130. These mechanisms require high-density molecular gas, in addition to strong near-IR or UV emission for the IR and chemical pumping, respectively. Although our analysis does not favor any of these mechanisms over the rest, the physical conditions of the gas are not similar to those expected in star-forming regions (PDRs). Consequently, the HF $J=1-0$ emission in NGC~7130 appears to be related to the AGN activity.

\section{Other Molecules}\label{s:other}

\subsection{Water}

We detected ortho- and para-water (o- and p-H$_2$O) transitions in four of our galaxies: two lines in UGC~05101, one in NGC~3227, four in NGC~7130, and one in NGC~7582 (see Table \ref{tab:lines}). All the transition are detected in emission. These water lines are weaker than the CO lines present in the \FTS\ spectra of these Seyfert galaxies. This is also the case in the starburst galaxy M~82 and the Seyfert 2 NGC~1068 (see \citealt{Kamenetzky2012, Spinoglio2012}). On the contrary, in the ULIRG Arp~220 water lines are more intense than the mid-$J$ CO lines \citep{Rangwala2011}.

To model the excitation of water using radiation transfer models we would need to detect more water transitions in each galaxy (see e.g., \citealt{Spinoglio2012,Gonzalez-Alfonso2012}); therefore we did not attempt this detailed analysis.

IR pumping is an efficient mechanism to excite water molecules. In particular the absorption of far-IR photons at 75.4\,\micron\ leads o-water molecules to the 3$_{21}$ energy level \citep{Gonzalez-Alfonso2010}. In UGC~05101 and NGC~7582, the o-H$_2$O $3_{21}-3_{12}$ line at 1163\,GHz is the brightest water line in the observed frequency range. Thus, in these two galaxies, IR pumping may be important for the excitation of water. On the contrary, in NGC~3227 and NGC~7130 this line is not detected, hence the contribution of IR pumping for water excitation in these galaxies could be lower.

\subsection{OH$^+$}\label{ss:OH}

The formation of OH$^+$ is in general associated to X-ray\slash cosmic-ray ionizations, although it is also formed in the transition region from ionized to molecular gas in PDRs (e.g., \citealt{Sternberg1995,Hollenbach2012}). This molecule is detected in the spectra of galaxies, in absorption in Arp~220 \citep{Rangwala2011,Gonzalez-Alfonso2012} and M~82 \citep{Kamenetzky2012}, and in emission in the Seyfert galaxies NGC~1068 \citep{Spinoglio2012} and Mrk~231 \citep{vanderWerf2010}. Its detection is interpreted as an indication of XDR excitation in these galaxies.

Three OH$^+$ lines are detected in emission in the \FTS\ spectrum of NGC~7130 suggesting the presence of an XDR in this galaxy. The hyperfine structure of these lines is not resolved with the spectral resolution of \FTS; therefore we measured the sum of the fluxes of the hyperfine transitions (Table \ref{tab:lines_uncommon}).

The Einstein coefficients of the hyperfine transitions were obtained from the Cologne database for molecular spectroscopy (CDMS; \citealt{Muller2005}). They were averaged for each fine structure level: $A_{ul} = \sum_{ij} g_i A_{ij}\slash \sum_i g_i$, where $i$ and $j$ correspond to each hyperfine upper and lower levels respectively. Then, the transition probabilities for the 909, 971, and 1033\,GHz OH$^+$ transitions are 1.57$\times$10$^{-2}$, 1.82$\times$10$^{-2}$, and 2.11$\times$10$^{-2}$\,s$^{-1}$, respectively.

Using Equation \ref{eqn:column_up} we obtained that the beam averaged column densities for NGC~7130 are 1.7$\times$10$^{11}$, 1.7$\times$10$^{11}$, and 1.6$\times$10$^{11}$\,cm$^{-2}$ for the upper levels of the 909, 971, and 1033\,GHz lines (assuming an 18\arcsec\ beam).

\section{Conclusions}\label{s:conclusions}

We presented the sub-millimeter spectra of eleven active galaxies observed with \FTS. Our study is focused on the spectral lines tracing the cold and warm molecular gas phases in these galaxies. Specifically the CO ladder from $J_{\rm up}=4$ to 13 lies in this spectral range, as well as the two fine structure [\ion{C}{1}] lines at 492 and 809\,GHz. For a few galaxies we also detected HF, H$_2$O, and OH$^{+}$ lines. In addition, we analyzed new IRAM 30\,m observations of the \CO1 and $J=2-1$ transitions for three of these galaxies.
The main results are summarized as follows.

\begin{enumerate}
\item We modeled the CO SLED (from $J_{\rm up}=1$ to $J_{\rm up}=7-12$) of six Seyfert galaxies in our sample using the escape probability approximation. For five of them the kinetic temperature and H$_2$ density of the warm molecular gas are similar ($n_{\rm H_2}\!\sim$10$^{3.2}$--10$^{3.9}$\,cm$^{-3}$ and $T_{\rm kin}\!\sim$300--800\,K). This warm molecular gas seems to be the same gas traced by the mid-IR H$_2$ S(1) 17.03\micron\ rotational line. The heating mechanism of the warm gas is not determined since
PDR, XDR, and shocks model can be combined to explain the observed CO SLEDs. However it is likely related to the star-formation activity since the $L_{\rm CO}$ is related to the IR luminosity but not to the AGN luminosity in these Seyfert galaxies.

\item We also used a radiative transfer model to interpret the [\ion{C}{1}] emission. The contribution from the warm molecular gas to the [\ion{C}{1}] emission seems to be small in these galaxies. Conversely, this emission can arise from cold ($T_{\rm kin}< 30$\,K) and dense ($n_{\rm H_2}>10^3$\,cm$^{-2}$) molecular gas where the [\ion{C}{1}] lines are thermalized. The C$^0$ abundance with respect to H$_2$ is 3$\times$10$^{-5}$--8$\times$10$^{-4}$; therefore the C$^0$\slash CO abundance ratio is 0.1--2, comparable to that observed in starburst galaxies. The \Cia\slash\CO1\ ratio ranges between 15 and 85 in our galaxies. It is consistent with that expected in PDRs, although for NGC~4388 XDR excitation could be possible.

\item In two galaxies we detected the HF $J=1-0$ transition at 1232\,GHz. In UGC~05101 it is detected in absorption, and implies a column density $N_{\rm HF} > (1.3 \pm 0.3)\times 10^{14}$\,cm$^{-2}$. In NGC~7130 the HF transition is observed in emission. We proposed three excitation mechanisms, near-IR pumping, chemical pumping, and electron collisions, that can produce the HF $J=1-0$ emission. From our analysis the three mechanisms seem to be plausible. They require very dense molecular gas, and, for the IR or chemical pumping, strong near-IR or UV fields. These conditions suggest that the HF emission in NGC~7130 is associated with the AGN activity.

\item We are not able to apply any radiative model to the water emission. However the detection of the o-H$_2$O $3_{21}-3_{12}$ transition at 1163\,GHz suggests that IR pumping (due to strong far-IR continuum) could play an important role in the excitation of water molecules in some of these galaxies.

\item In NGC~7130 we detected three OH$^+$ lines in emission (1$_0$--0$_1$ at 909\,GHz, 1$_2$--0$_1$ at 972\,Ghz, and 1$_1$--0$_1$ at 1033\,GHz). We derived a column density of $\sim$1.7$\times$10$^{11}$\,cm$^{-2}$ for the upper levels of these three transitions. These detections suggest the presence of an XDR in NGC~7130.
\end{enumerate}

\acknowledgments

We thank Rui-Qing Mao for kindly providing their published \CO3 data for NGC~3227 and NGC~3982. We are grateful to N. Sacchi, D. Truco and the IRAM 30\,m staff for their support during the observations.
We thank the referee for comments that improved the paper. This work has been funded by the Agenzia Spaziale Italiana (ASI) under contract I/005/11/0.

SPIRE has been developed by a consortium of institutes led by Cardiff Univ. (UK) and including: Univ. Lethbridge (Canada); NAOC (China); CEA, LAM (France); IFSI, Univ. Padua (Italy); IAC (Spain); Stockholm Observatory (Sweden); Imperial College London, RAL, UCL-MSSL, UKATC, Univ. Sussex (UK); and Caltech, JPL, NHSC, Univ. Colorado (USA). This development has been supported by national funding agencies: CSA (Canada); NAOC (China); CEA, CNES, CNRS (France); ASI (Italy); MCINN (Spain); SNSB (Sweden); STFC, UKSA (UK); and NASA (USA).
This research has made use of the NASA/IPAC Extragalactic Database (NED) which is operated by the Jet Propulsion Laboratory, California Institute of Technology, under contract with the National Aeronautics and Space Administration.

\end{document}